# Description of Excitations in Odd Nonmagic Nuclei by the Green's Function Method

**A. V. Avdeenkov and S. P. Kamerdzhiev**

*Institute of Physics and Power Engineering, Obninsk, Kaluga oblast, 249020 Russia*

**Abstract**—General equations for single-particle Green's functions in nonmagic nuclei have been derived. A pairing mechanism similar to the Bardeen–Cooper–Schrieffer mechanism is singled out explicitly in these equations. A refining procedure for phenomenological single-particle energies and for the gap has been developed to avoid doubly taking into account mixing with phonons for the situation in which the input data for the problem in question are formulated in terms of these phenomenological quantities. The resulting general equations are written within the second-order approximation in the phonon-creation amplitude. This corresponds to taking into account quasiparticle⊗phonon configurations and is shown to be a fairly good approximation for semimagic nuclei. A secular equation for calculating excitations in odd nuclei that takes fully into account ground-state correlations and which is invariant under the reversal of the sign of the energy variable has been derived in this approximation. Distributions of single-particle strengths have been computed for $^{119}$Sn and $^{121}$Sn. Reasonably good agreement with available experimental data has been obtained.

## 1. INTRODUCTION

To a considerable extent, successes of the modern microscopic theory of nuclear structure are due to taking into account quasiparticle–phonon interaction both in magic and in nonmagic nuclei (see [1–10] and references therein). For even–even nuclei, this involves going beyond the random-phase approximation (RPA and QRPA for magic and nonmagic nuclei, respectively)—that is, allowing for configurations more complicated than those considered in the RPA or QRPA (above all, $1p1h$⊗phonon configurations). For odd nuclei, it proved necessary to take into account at least $1p$⊗phonon or $1h$⊗phonon configurations. It was shown in [2, 5, 6, 9, 10] for even–even nuclei and in [2, 3, 6] for odd nuclei that the inclusion of quasiparticle–phonon interaction is of paramount importance for explaining qualitatively and quantitatively many observable effects.

The main distinction of the theory of nonmagic nuclei from the theory of magic nuclei is that, in the former case, it is necessary to take into account Cooper pairing in the nuclear ground state. In order to describe excited states at the QRPA level, it is necessary to allow for not only particle–hole ($ph$) configurations but also hole–hole ($hh$) and particle–particle ($pp$) configurations and their superpositions. The situation is often described in terms of gap variations in an external field [11] or in terms of the inclusion of the $pp$ channel and pair phonons.

Within the Green's function method, the procedure for taking into account the above complicated configurations in magic and (mag ± 1 particle) nuclei consists in explicitly isolating, in the full mass operator, the simplest pole term involving $g^2$, where $g$ is the amplitude of the production of a lowly lying collective phonon [7, 8, 12],

$M =$ 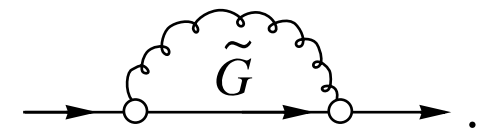 . (1)

(In such nuclei, the $g^2$ approximation is quite satisfactory [4, 7].) For (mag ± 1 particle) nuclei, this makes it possible to calculate the features of excited states representing superpositions of the form $1p$ + $1p$⊗phonon [12]. For doubly magic nuclei, the isolation of the term in (1) corresponds to taking into account RPA + $1p1h$⊗phonon configurations—in other words, this means that, in the irreducible $ph$ amplitude, which appears in the input RPA equation written in terms of Green's functions [11], we single out the most dangerous terms as represented by graphs with insertions and with a transverse phonon [7, 8]. These terms describe, respectively, the complication of single-particle motion due to mixing with phonons and the emergence of the phonon-exchange-induced retarded interaction in addition to the conventional $ph$ interaction (for details, see, for example, [8]). A further development and applications of this approach resulted in a successful description of giant resonances in magic nuclei; in particular, this made it possible to take consistently into account ground-state correlations and a single-particle continuum and to implement numerically the entire approach for a nonseparable effective interaction [10].

We will see below that the $g^2$ approximation is quite satisfactory for nuclei that are magic in one nucleon species; therefore, it is advisable to implement the above program for such nuclei. To be more specific, we deem it appropriate to consider explicitly, in addition to (1), pole terms in the anomalous mass operators; that is,

$M^{(1)} =$ 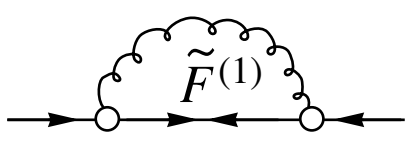, $M^{(2)} =$ 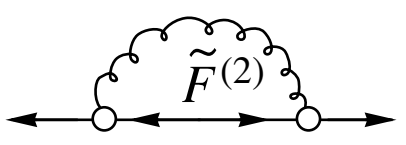 . (2)

In (1) and (2) and in the ensuing analysis, pair phonons are disregarded, since their contribution is expected to be small.

For even–even nuclei, this will lead, by analogy with what was said above, to the emergence of the corresponding graphs featuring insertions and a transverse phonon. Since these equations are very complicated, their analysis will be performed later. To begin with, it is appropriate to consider a simpler and more pressing question of whether it is possible to implement the Green's function method for a specific example in odd nuclei that corresponds to taking into account quasiparticle–phonon interaction in the simplest $g^2$ approximation for the mass operator. This is the prime objective of the present study.

Since the microscopic theory of Bardeen–Cooper–Schrieffer (BCS) pairing considers *pp* and *hh* channels, the addition of the pole terms $M^{(1)}$ and $M^{(2)}$ (2) to the conventional energy-independent mass operators

$\Delta^{(1)} \equiv$ 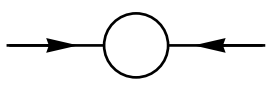, $\Delta^{(2)} \equiv$ 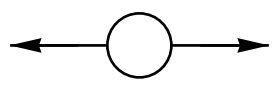, (3)

which correspond to the BCS approach [11], implies the inclusion of the quasiparticle–phonon pairing mechanism, which is usually disregarded in the microscopic theory of pairing in nuclei. In the limit of weak quasiparticle–phonon interaction ($g^2 \ll 1$), this mechanism, which was proposed by Eliashberg in the theory of superconductivity [13], reduces to the BCS mechanism dominated by electron–electron interaction in solids. In the case of pairing in nuclei, the BCS equation involves the effective *pp* interaction, which is usually determined from a comparison with experimental data. It can therefore be said that, in nuclei, the quasiparticle–phonon pairing mechanism is usually taken into account to an extent that can be absorbed by the BCS mechanism. In a perfect analogy with the *ph* channel [7, 8], this means that, when the quasiparticle–phonon interaction is included explicitly, the original phenomenological parameters of the effective *pp* interaction must be modified (refined) in order to avoid doubly taking into account quasiparticle–phonon interaction in the *pp* channel. A simpler way to achieve this consists in developing and implementing a phenomenological procedure that is ideologically similar to the procedure of refining single-particle levels in magic nuclei [7–10]. In doing this, it proved necessary to apply the procedure for extracting single-particle energies from the observed quasiparticle energies of nonmagic nuclei. Thus, the second objective of the present study is to evolve and implement the aforementioned two procedures. Finally, the third objective is to apply the proposed approach to calculating the distributions of single-particle strengths in some odd nonmagic nuclei.

The exposition is organized as follows. Section 2 is devoted to deriving general relations that are necessary for solving our problem featuring two pairing mechanisms. In Section 3, all the results are obtained for the realistic case of the $g^2$ approximation; in particular, a comparison is made with the corresponding version of the quasiparticle–phonon model from [1, 2]. In Section 4, the developed approach is used to compute the distributions of single-particle strengths in $^{119}$Sn and $^{121}$Sn nuclei.

## 2. GENERAL RELATIONS FOR NONMAGIC NUCLEI

### *2.1. Equations for Single-Particle Green's Functions in Nonmagic Nuclei*

In nonmagic nuclei, two anomalous Green's functions $F^{(1)}$ and $F^{(2)}$ associated with the Bose condensate of Cooper pairs [11] are used in addition to the conventional (causal) Green's functions $G_C$ and $G_C^{(h)}$. If only the mean (self-consistent) field and nucleon–nucleon interaction leading to a pairing gap, which is independent of energy and which usually obeys the BCS equation, are taken into account, all these Green's functions are determined by the set of Gor'kov equations, whose solutions are well known both for infinite [14] and for finite [11] Fermi systems. But if it is necessary to take into account an additional interaction—this is the quasiparticle–phonon interaction in our case—it is convenient to obtain equations for the single-particle Green's functions in a form where the known Gor'kov Green's functions appear to be the free Green's functions representing the zeroth-approximation solution to the problem in question. In terms of the Hamiltonian formalism, this corresponds to going over to Bogolyubov quasiparticles. It will be shown below that, if the phenomenological mean field, the *pp* interaction, and the corresponding BCS gap are taken to be the input data for the problem, then they must be modified (refined) in order to avoid doubly taking into account the quasiparticle–phonon interaction.

Explicitly, we will consider only two Green's functions in an *N*-particle Fermi system (the equations for the remaining two Green's functions are obtained in a similar way). In accordance with [11, 14], we set

$$\begin{aligned} G_C &= -i\langle N|T\hat{\psi}(x)\hat{\psi}^+(x')|N\rangle, \\ F^{(2)} &= \langle N+2|T\hat{\psi}^+(x)\hat{\psi}^+(x')|N\rangle, \end{aligned} \tag{4}$$

where the subscript C, which is retained only in this section, indicates that we take into account general Cooper pairing, which can be described either in terms of BCS theory or with allowance for the quasiparticle–phonon interaction.

The Green's functions in (4) satisfy the set of equations

$$\begin{aligned} G_C &= G_0 + G_0\Sigma G_C - G_0\Sigma^{(1)}F^{(2)}, \\ F^{(2)} &= G_0^{(h)}\Sigma^{(h)}F^{(2)} + G_0^{(h)}\Sigma^{(2)}G_C, \end{aligned} \tag{5}$$

which generalize the Dyson equation to the case of pairing. In equations (5), $\Sigma$, $\Sigma^{(h)}$, $\Sigma^{(1)}$, and $\Sigma^{(2)}$ are the corresponding total irreducible self-energy parts (mass operators), while $G_0$ and $G_0^{(h)}$ are the free Green's functions (that is, the Green's functions of an ideal Fermi gas). Obviously, there do not exist free functions $F_0^{(2)}$ and $F_0^{(1)}$. Throughout this section, we employ the condensed symbolic notation, which can easily be expanded [11, 14].

In principle, equations (5) contain the entire body of information about an odd Fermi system. In order to obtain treatable equations, it is necessary to isolate explicitly the well-known blocks—that is, the mean field and Cooper pairing described by a BCS-type equation. For this, each of the total mass operators in the system of equations (5) is represented as the sum of two terms. Of these, one is independent of energy, while the other is energy-dependent, but it is not specified for the time being. Accordingly, we have

$$\begin{aligned} \Sigma(\varepsilon) &= \tilde{\Sigma} + M(\varepsilon), \quad \Sigma^{(h)}(\varepsilon) = \tilde{\Sigma}^{(h)} + M^{(h)}(\varepsilon), \\ \Sigma^{(1)}(\varepsilon) &= \tilde{\Sigma}^{(1)} + M^{(1)}(\varepsilon), \quad \Sigma^{(2)}(\varepsilon) = \tilde{\Sigma}^{(2)} + M^{(2)}(\varepsilon), \end{aligned} \tag{6}$$

where $\tilde{\Sigma}$ and $\tilde{\Sigma}^{(h)}$ correspond to the mean field, while $\tilde{\Sigma}^{(1)}$ and $\tilde{\Sigma}^{(2)}$ are associated with pairing described by a mechanism of the BCS type.

We introduce the Green's functions

$$\tilde{G} = G_0 + G_0\tilde{\Sigma}\tilde{G}, \quad \tilde{G}^{(h)} = G_0^{(h)} + G_0^{(h)}\tilde{\Sigma}^{(h)}\tilde{G}^{(h)}, \tag{7}$$

which determine refined quasiparticles in the mean field (see below). The system of equations (5) can then easily be recast into the form

$$\begin{aligned} G_C &= \tilde{G} + \tilde{G}MG_C - \tilde{G}\Sigma^{(1)}F^{(2)}, \\ F^{(2)} &= \tilde{G}^{(h)}M^{(h)}F^{(2)} + \tilde{G}^{(h)}\Sigma^{(2)}G_C. \end{aligned} \tag{8}$$

Formally, these equations coincide with those that Eliashberg [13] solved for an infinite system by approximating the quantities $M$, $M^{(h)}$, $\Sigma^{(1)}$, and $\Sigma^{(2)}$ by nonlinear mass operators of the type (14) (see below). The situation for nuclei is, however, different: the gap can be determined from experimental data or by solving the BCS equation with an interaction potential chosen phenomenologically. Instead of solving equations (8) featuring an unknown refined *pp* interaction, we can therefore rely on the fact that there exist a known mean field and a known gap ($\varepsilon_\lambda$ and $\Delta_\lambda$). The input parameters of the problem will then be the quantities $\tilde{\varepsilon}_\lambda$ and $\tilde{\Delta}_\lambda$ corresponding to the mean field $\tilde{\Sigma}$ and BCS-type pairing, and we can further take into account the quasiparticle–phonon interaction in the $g^2$ approximation (see Section 3 below).

In what follows, we everywhere imply that the "observed" mean field described by a phenomenological potential (usually of the Woods–Saxon type) and the observed gap extracted from experimental data or determined as a solution to the BCS equation with a *pp* interaction chosen phenomenologically are the input known values. We also take into account the terms $M$, $M^{(h)}$, $M^{(1)}$, and $M^{(2)}$ in the mass operator (for the sake of brevity, we denote them by $M^i$), which obviously contribute to the aforementioned phenomenological quantities inasmuch as they can be reduced to energy-independent functions. In order to avoid doubly taking into account the quantities $M^i$, we must eliminate them from the phenomenological quantities—that is, to refine them. These refined terms will be tilde-labeled. The procedure for refining single-particle energies that results in constructing a new single-particle basis for magic nuclei was used in [7–10]. Here, we refine the gap to determine the quantities $\tilde{\Delta}^{(1)}$ and $\tilde{\Delta}^{(2)}$. This will be done in Subsection 3.3.

Using representation (6) and following [15], we can rewrite equations (8) as[1)]

$$\begin{aligned} G_C &= \tilde{G}_C + \tilde{G}_C M G_C - \tilde{F}^{(1)}M^{(h)}F^{(2)} \\ &\quad - \tilde{G}_C M^{(1)}F^{(2)} - \tilde{F}^{(1)}M^{(2)}G_C, \\ F^{(2)} &= \tilde{F}^{(2)} + \tilde{F}^{(2)}MG_C + \tilde{G}_C^{(h)}M^{(h)}F^{(2)} \\ &\quad - \tilde{F}^{(2)}M^{(1)}F^{(2)} + \tilde{G}_C^{(h)}M^{(2)}G_C. \end{aligned} \tag{9}$$

In the graphical form, this set of equations is represented as

[1)] Without refining the phenomenological quantities in the sense explained above, equations (9) can also be obtained from a general set of equations in a matrix form that are presented in [16]. One of the present authors (S.P.K.) is grateful to V.I. Tselyaev for pointing out this reference.

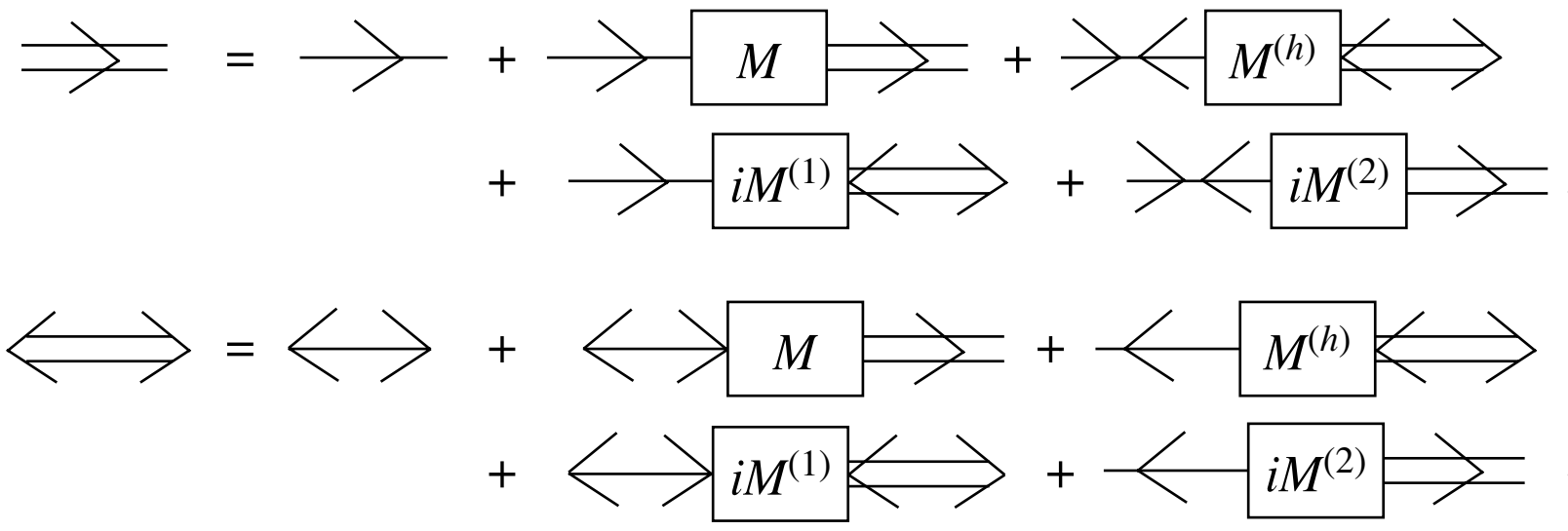

where rectangular blocks stand for the mass operators $M^i$, which we do not specify for the time being. In deriving (9), we used the equations (for details, see [15])

$$\tilde G_C = \tilde G - \tilde G\tilde\Delta^{(1)}\tilde F^{(2)}, \quad \tilde G_C^{(h)} = \tilde G^{(h)} - \tilde G^{(h)}\tilde\Delta^{(2)}\tilde F^{(1)},$$
$$\tilde F^{(2)} = \tilde G^{(h)}\tilde\Delta^{(2)}\tilde G_C, \quad \tilde F^{(1)} = \tilde G\tilde\Delta^{(1)}\tilde G_C^{(h)}, \qquad (10)$$

which determine the Gor'kov Green's functions and which involve the refined quantities $\tilde\varepsilon_\lambda$ and $\tilde\Delta_\lambda$.

The Green's functions $\tilde G$ and $\tilde G^{(h)}$ appearing in (10) satisfy equations (7) and determine the mean field with single-particle energies $\tilde\varepsilon_\lambda$ and functions $\tilde\phi_\lambda$; that is, they specify a new, refined basis. All the relations that appear below are written in this basis.

The BCS pairing mechanism—more precisely, its analog that differs from the conventional one by refining—was introduced in (10) through the mass operator [11]

$$\tilde\Sigma_{BCS} = -\tilde\Delta^{(1)}\tilde G^{(h)}\tilde\Delta^{(2)} \qquad (11)$$

and the Green's function

$$\tilde G_C = \tilde G + \tilde G\tilde\Sigma_{BCS}\tilde G_C. \qquad (12)$$

By solving the set of equations (10) [11], we find that the "bare" Green's functions in the set of equations (9) are given by

$$\tilde G_{C\lambda}(\varepsilon) = \tilde G_{C\lambda}^{(h)}(-\varepsilon) = \frac{u_\lambda^2}{\varepsilon - \tilde E_\lambda + i\delta} + \frac{v_\lambda^2}{\varepsilon + \tilde E_\lambda - i\delta}, \qquad (13)$$

$$\tilde F_\lambda^{(1)} = \tilde F_\lambda^{(2)} = -\frac{\tilde\Delta_\lambda}{2\tilde E_\lambda}\left(\frac{1}{\varepsilon - \tilde E_\lambda + i\delta} - \frac{1}{\varepsilon + \tilde E_\lambda - i\delta}\right),$$

where $u_\lambda^2 = 1 - v_\lambda^2 = (\tilde E_\lambda + \tilde\varepsilon_\lambda)/(2\tilde E_\lambda)$ and $\tilde E_\lambda = \sqrt{\tilde\varepsilon_\lambda^2 + \tilde\Delta_\lambda^2}$.

Thus, equations (9) are adequate to the problem in question. The refined quantities $\tilde\varepsilon_\lambda$ and $\tilde\Delta_\lambda$ will be defined below. From the set of equations (9), we can obtain useful approximations by specifying mass operators $M^i$. For example, we can generalize the extensively used nonlinear approximation for magic nuclei that corresponds to the diagram

$M =$ 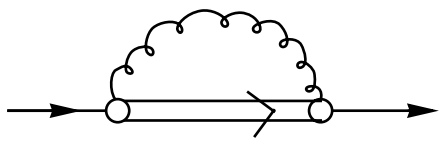 , (14)

where the double line represents the total Green's function. This nonlinear $g^2$ approximation means that the total phonon Green's function and the vertex that must appear in the mass operator are taken in the simplest form. In the case considered here, this corresponds to the approximation where, in each operator $M^i$, we take one term involving a particle–hole phonon (more precisely, the particle–hole component of the phonon determined by solving the full system of QRPA equations). As a result, we arrive at a nonlinear set of four equations. The first two of these can be graphically represented as

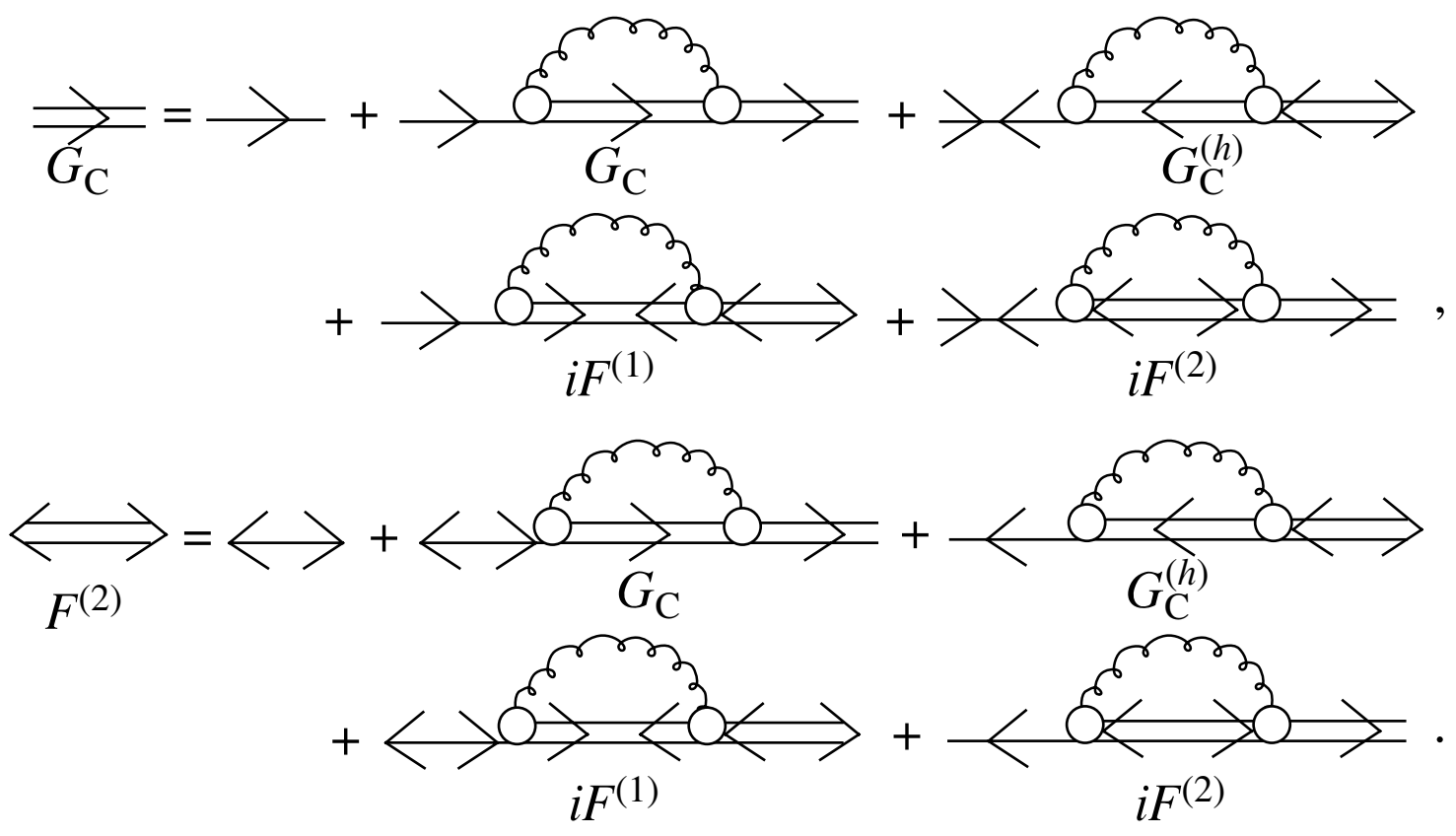

(15)

The equations for $G_C^{(h)}$ and $F^{(1)}$ have a similar structure. It is not clear whether this approximation is legitimate; moreover, it is very difficult to solve this set of equations. In view of this and the fact that the $g^2$ approximation is fairly reasonable for nuclei with one filled shell (see Subsection 3.1), it is advisable to go over to the simpler case of the pure $g^2$ approximation by linearizing the set of equations (15). Prior to doing this, we will derive some useful relations for the Green's functions and mass operators from the general equations (5).

### *2.2. General Relations for Green's Functions and Mass Operators: Refining Single-Particle Energies and the Gap*

Let us introduce the Green's functions

$$\begin{aligned} G &= G_0 + G_0(\tilde{\Sigma} + M)G = \tilde{G} + \tilde{G}MG, \\ G^{(h)} &= G_0^{(h)} + G_0^{(h)}(\tilde{\Sigma}^{(h)} + M^{(h)})G^{(h)} \\ &= \tilde{G}^{(h)} + \tilde{G}^{(h)}M^{(h)}G^{(h)}, \end{aligned} \tag{16}$$

where the Green's functions $\tilde{G}$ and $\tilde{G}^{(h)}$ are defined in (7). The set of equations (5) can then be recast into the form

$$\begin{aligned} G_C &= G - G\Sigma^{(1)}F^{(2)}, \\ F^{(2)} &= G^{(h)}\Sigma^{(2)}G_C. \end{aligned} \tag{17}$$

Substituting $F^{(2)}$ into the expression for $G_C$, we obtain

$$G_C = G - G\Sigma^{(1)}G^{(h)}\Sigma^{(2)}G_C \equiv G + GM_CG_C, \tag{18}$$

where the mass operator

$$M_C = -\Sigma^{(1)}G^{(h)}\Sigma^{(2)} \tag{19}$$

generalizes expression (11), which determines conventional BCS pairing. As a matter of fact, the expression in (19) is a general representation for the mass operator responsible for pairing in the Fermi system.

Equation (18) can easily be reduced to the form

$$G_C = \tilde{G} + \tilde{G}(M + M_C)G_C. \tag{20}$$

With the aid of the above results, the relations obtained in the theory of finite Fermi systems between single-particle energies and the mass operator [11] can be generalized with allowance for pairing and quasiparticle–phonon interaction. For the sake of simplicity—and in view of the fact that, in the ensuing applications, we will need only this case—all the quantities in (18) will be considered in the diagonal approximation in the single-particle subscript $\lambda$:

$$G_{C\lambda} = G_\lambda - G_\lambda\Sigma_\lambda^{(1)}G_\lambda^{(h)}\Sigma_\lambda^{(2)}G_{C\lambda}. \tag{21}$$

Following (17), we further represent the mass operators $M$ and $M^{(h)}$ as the sum of components that are even and odd in energy—for example, we set

$$M = M_{\text{even}} + M_{\text{odd}}. \tag{22}$$

Defining the sought energies as the poles of the Green's function $G_{C\lambda}$ and using (21), we then find that the formal expression for these energies is given by

$$E_{\lambda\eta} = \sqrt{\varepsilon_{\lambda\eta}^2 + \Delta_{\lambda\eta}^2}, \tag{23}$$

where

$$\varepsilon_{\lambda\eta} = \frac{\tilde{\varepsilon}_\lambda + M_{\text{even}\lambda}(E_{\lambda\eta})}{1 + q_{\lambda\eta}}, \tag{24}$$

$$\Delta_{\lambda\eta}^2 = \frac{\Sigma_\lambda^{(1)}(E_{\lambda\eta})\Sigma_\lambda^{(2)}(E_{\lambda\eta})}{(1 + q_{\lambda\eta})^2}. \tag{25}$$

Here, we introduced the notation

$$q_{\lambda\eta} = -\frac{M_{\text{odd}\lambda}(E_{\lambda\eta})}{E_{\lambda\eta}}.$$

The subscript $\eta$ numbers solutions to the set of equations (23)–(25). The difference between what was done here and the approach used in [17] is that we avoid doubly taking into account the mass operators $M^i$ by introducing the unobservable refined quantities $\tilde{\varepsilon}_\lambda$ and $\tilde{\Delta}_\lambda$. Since the experimental single-quasiparticle energies used here must correspond to dominant levels (that is, those that have the maximum spectroscopic factor), the refining must be such that, upon solving the Dyson equation, one of the solutions coincides with the experimental value and that the level remains dominant. In other words, these experimental single-quasiparticle energies serve as input data for the entire problem. The single-particle energies in question and the gap values corresponding to these energies will be denoted by $\varepsilon_\lambda$ and $\Delta_\lambda$. Proceeding from the above condition and using (6), (24), and (25), we find that the observed and refined values are related by the equations

$$\begin{aligned} \varepsilon_\lambda &= \frac{\tilde{\varepsilon}_\lambda + M_{\text{even}\lambda}(E_\lambda)}{1 + q_\lambda(E_\lambda)}, \\ \Delta_\lambda \equiv \Delta_\lambda^{(1,2)} &= \frac{\tilde{\Delta}_\lambda^{(1,2)} + M^{(1,2)}(E_\lambda)}{1 + q_\lambda(E_\lambda)}, \end{aligned} \tag{26}$$

where $E_\lambda = \sqrt{\varepsilon_\lambda^2 + \Delta_\lambda^2}$. In equations (23)–(26), the energies $\tilde{\varepsilon}_\lambda$ and $\varepsilon_\lambda$ are reckoned from the corresponding chemical potentials $\mu$ and $\tilde{\mu}$.

From equation (26), it can easily be found that refining in magic nuclei is performed according to the relation [7–10, 12]

$$\varepsilon_\lambda = \tilde{\varepsilon}_\lambda + M_\lambda(\varepsilon_\lambda). \tag{27}$$

## 3. $g^2$ APPROXIMATION FOR NONMAGIC NUCLEI

It can be seen from the above discussion that, in the nonlinear $g^2$ approximation, the problem of calculating the features of odd nonmagic nuclei is very difficult, especially as it is necessary to implement the nonlinear procedure of refining according to (26). It is therefore advisable to begin by considering the simpler problem for odd nuclei in the linearized $g^2$ approximation.

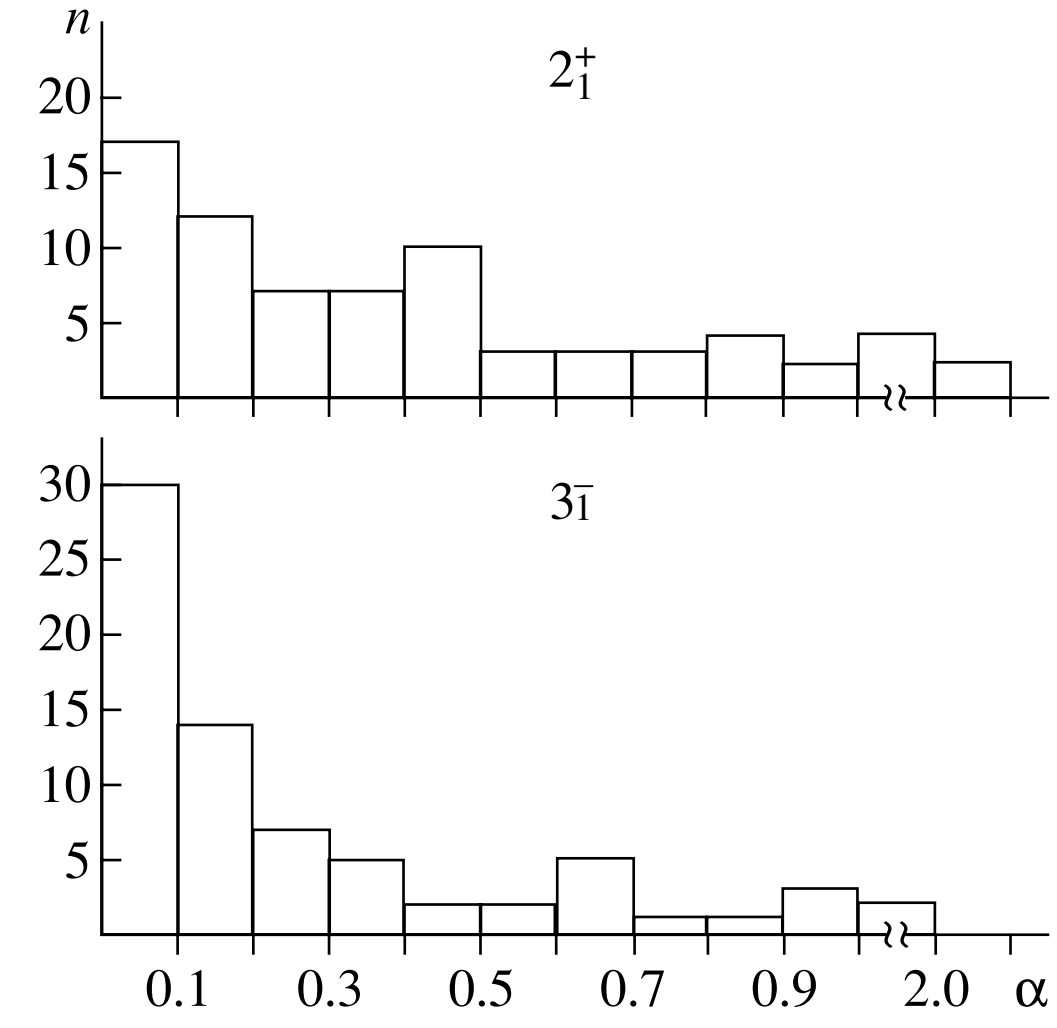

**Fig. 1.** Histograms representing the distributions of the parameter α (28) for the $2_1^+$ and $3_1^-$ phonons in $^{120}$Sn ($n$ is the number of α values in a given interval).

### *3.1. Assessing the Applicability of the $g^2$ Approximation*

It was shown in [4] that, for doubly magic nuclei, the dimensionless coupling constant $g^2$ satisfies the condition $g^2 < 1$ for all lowly lying phonons. For nonmagic nuclei, which involve a lowly lying collective $2_1^+$ phonon, $g^2$ can be on the order of unity [4]. It is interesting to assess this quantity for nuclei that are magic only in one nucleon species, in which case the degree of collectivization of the $2_1^+$ level is not very high.

Figure 1 shows histograms representing the distribution of the parameter [4]

$$\alpha = \frac{\langle 1\|g\|2\rangle^2}{(2j_1+1)\omega_s^2} \qquad (28)$$

for the $^{120}$Sn nucleus. Here, $\langle 1\|g\|2\rangle$ (1 represents the set of single-particle quantum numbers $n_1$, $l_1$, and $j_1$) are the reduced matrix elements of the amplitude of the production of a phonon with energy $\omega_s$ that were calculated in the present study on the basis of the theory of finite Fermi systems (see Subsection 4.1). It can be seen that, by and large, the requirement $\alpha < 1$ is satisfied. But our problem involves the slightly different quantity (see Appendix)

$$\alpha' = \frac{\langle 1\|g\|2\rangle^2}{(2j_1+1)(E_1^2-(E_2+\omega_s)^2)}, \qquad (29)$$

where $E_1$ and $E_2$ are the energies of Bogolyubov quasiparticles. The calculations revealed that this parameter is several times as small as α. Thus, the approximation of $g^2 < 1$ is quite reasonable in our problem.

### *3.2. Equation for the Green's functions in the $g^2$ Approximation and Secular Equation*

Linearizing the set of four equations (15), we arrive at equations for two Green's functions. Diagrammatically, they are represented as (hereafter, the subscript C is suppressed)

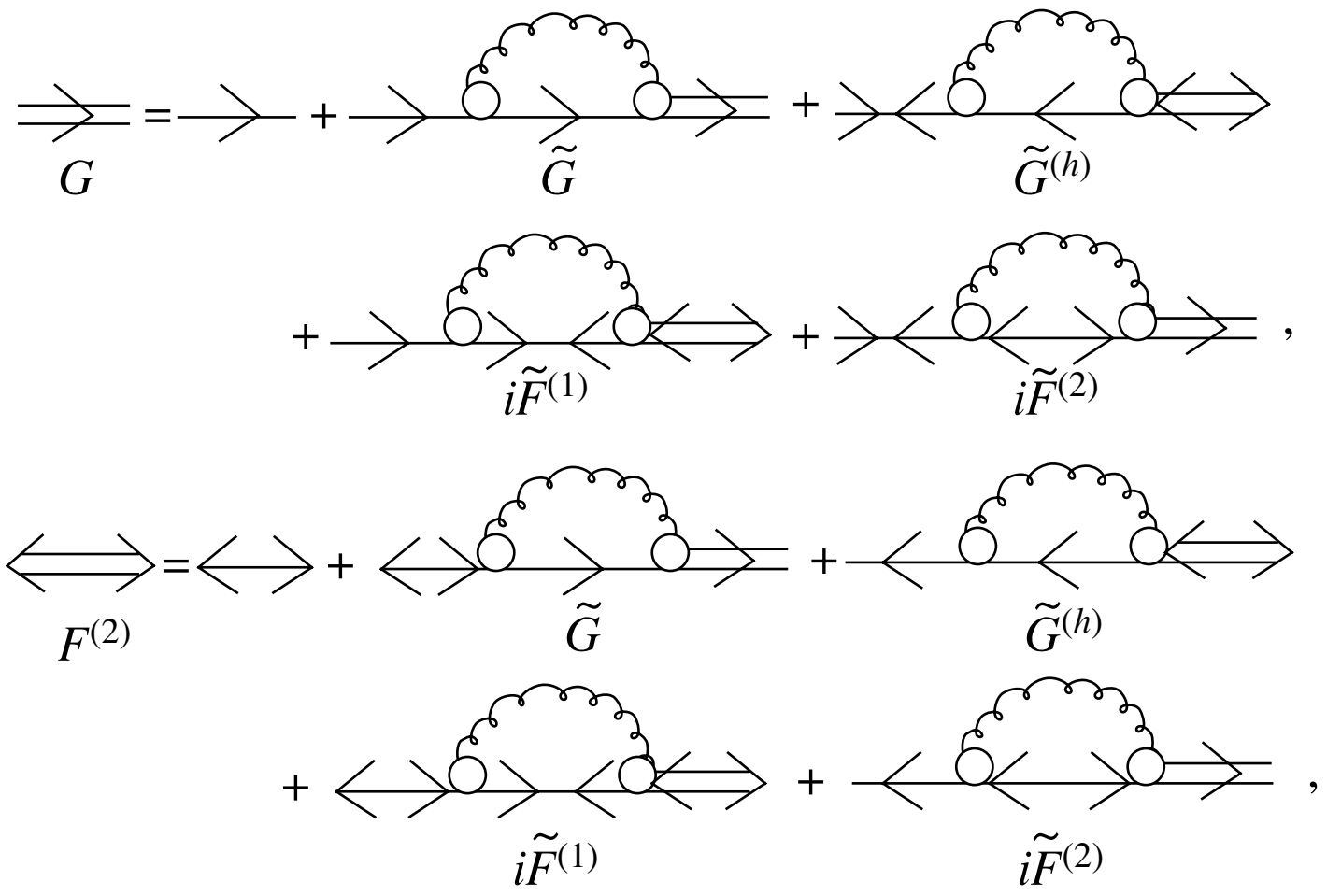

(30)

where the bare Green's functions $\tilde{G}$, $\tilde{G}^{(h)}$, $\tilde{F}^{(1)}$, and $\tilde{F}^{(2)}$ are given by (13), while the expressions for the mass operators are presented in the Appendix. In the diagonal approximation with respect to the subscript λ (as was indicated above, this approximation is quite reasonable in the problem at hand), equations (30) reduce to

$$\begin{aligned} G_\lambda &= \tilde{G}_\lambda + \tilde{G}_\lambda M_\lambda G_\lambda - \tilde{F}_\lambda^{(1)} M_\lambda^{(h)} F_\lambda^{(2)} \\ &\quad - \tilde{G}_\lambda M_\lambda^{(1)} F_\lambda^{(2)} - \tilde{F}_\lambda^{(1)} M_\lambda^{(2)} G_\lambda, \\ F_\lambda^{(2)} &= \tilde{F}_\lambda^{(2)} + \tilde{F}_\lambda^{(2)} M_\lambda G_\lambda + \tilde{G}_\lambda^{(h)} M_\lambda^{(h)} F_\lambda^{(2)} \\ &\quad - \tilde{F}_\lambda^{(2)} M_\lambda^{(1)} F_\lambda^{(2)} + \tilde{G}_\lambda^{(h)} M_\lambda^{(2)} G_\lambda. \end{aligned} \quad (31)$$

From (31), we find that the secular equation for determining the energies of excited states in an odd nucleus has the form

$$\Xi_\lambda(\varepsilon) \quad (32)$$

$$\equiv \begin{pmatrix} 1 - \tilde{G}_\lambda M_\lambda + \tilde{F}_\lambda^{(1)} M_\lambda^{(2)} & \tilde{F}_\lambda^{(1)} M_\lambda^{(h)} + \tilde{G}_\lambda M_\lambda^{(1)} \\ -\tilde{F}_\lambda^{(2)} M_\lambda - \tilde{G}_\lambda^{(h)} M_\lambda^{(2)} & 1 - \tilde{G}_\lambda^{(h)} M_\lambda^{(h)} + \tilde{F}_\lambda^{(2)} M_\lambda^{(1)} \end{pmatrix} = 0.$$

Solving the set of equations (31), we arrive at

$$\begin{aligned} G_\lambda(\varepsilon) &= \frac{\varepsilon + \tilde{\varepsilon}_\lambda + M_\lambda^{(h)}(\varepsilon)}{\theta_\lambda(\varepsilon)}, \\ F_\lambda^{(2)}(\varepsilon) &= \frac{\tilde{\Delta}_\lambda + M_\lambda^{(2)}(\varepsilon)}{\theta_\lambda(\varepsilon)}, \end{aligned} \quad (33)$$

where

$$\begin{aligned} \theta_\lambda(\varepsilon) &= (\varepsilon - \tilde{\varepsilon}_\lambda - M_\lambda(\varepsilon))(\varepsilon + \tilde{\varepsilon}_\lambda + M_\lambda^{(h)}(\varepsilon)) \\ &\quad - (\tilde{\Delta}_\lambda + M_\lambda^{(1)}(\varepsilon))^2. \end{aligned}$$

The residues of the Green's function $G_\lambda$ determine the strengths of transitions to corresponding excited states (labeled with the subscript $\eta$). We have

$$S_{\lambda\eta}^{\pm} = \frac{(1 + q_{\lambda\eta})(E_{\lambda\eta} \pm \varepsilon_{\lambda\eta})}{\dot{\theta}_\lambda(E_{\lambda\eta})}, \quad (34)$$

where the overdot denotes differentiation with respect to energy.

Relations (32) and (34) determine the features of odd nonmagic nuclei. These relations can be improved by using more accurate expressions from [11] for the single-particle Green's functions describing odd nuclei. This is especially important for near-magic nuclei.

We further make the following approximations:

(i) The determinant in (32) will be used in the $g^2$ approximation; that is,

$$1 - \tilde{G}_\lambda M_\lambda + \tilde{F}_\lambda^{(1)} M_\lambda^{(2)} - \tilde{G}_\lambda^{(h)} M_\lambda^{(h)} + \tilde{F}_\lambda^{(2)} M_\lambda^{(1)} = 0. \quad (35)$$

(ii) In (35), we will not take into account refining, setting $\tilde{\varepsilon}_\lambda = \varepsilon_\lambda$ and $\tilde{\Delta}_\lambda = \Delta_\lambda$.

Upon the substitution of (13) and of the expressions from the Appendix, the secular equation takes the form

$$\begin{aligned} 1 &- \left\{ \frac{1}{\varepsilon - E_1} \sum_{s,2} g_{12}^2 \left( \frac{v_{12}^2}{\varepsilon - E_{2s}} + \frac{u_{12}^2}{\varepsilon + E_{2s}} \right) \right\} \\ &- \left\{ \frac{1}{\varepsilon + E_1} \sum_{s,2} g_{12}^2 \left( \frac{u_{12}^2}{\varepsilon - E_{2s}} + \frac{v_{12}^2}{\varepsilon + E_{2s}} \right) \right\} = 0, \end{aligned} \quad (36)$$

where $E_{2s} = E_2 + \omega_s$; the subscript 1 stands for the set of single-particle quantum numbers; the subscript $s$ represents all phonon quantum numbers; and

$$\begin{aligned} v_{12}^2 &= u_1^2 u_2^2 + v_1^2 v_2^2 - \frac{\Delta_1 \Delta_2}{2E_1 E_2}, \\ u_{12}^2 &= u_1^2 v_2^2 + u_2^2 v_1^2 + \frac{\Delta_1 \Delta_2}{2E_1 E_2}. \end{aligned} \quad (37)$$

Equation (36) without terms in the second braces corresponds to taking into account only the first (pole) terms in the Green's functions $\tilde{G}$, $\tilde{G}^{(h)}$, and $\tilde{F}^{(1)}$ in (13). This equation was obtained in the quasiparticle–phonon model (see [1]), but neither two-phonon components nor the Pauli exclusion principle was taken into account there. It allows only partly for ground-state correlations in an even–even nucleus. Corresponding corrections within the quasiparticle–phonon model were briefly considered in [18]. Our equations (32) and (36) are invariant under the substitution of $-\varepsilon$ for $\varepsilon$, and this is physically reasonable because of invariance under time reversal. This invariance is ensured by the terms in the second braces in (36).

### *3.3. Equation for the Gap and Refinement of the Gap*

The problem of taking into account quasiparticle–phonon interaction in the *pp* channel within the Green's function method was considered in the pioneering studies reported in [19] and, in a narrower context (for pairing in nuclei), in [20], where it was concluded that pairing in nuclei is entirely due to quasiparticle–phonon interaction and that it is a purely surface effect, at least in (mag ± 2) nuclei. However, more detailed investigations reported in [21], which were based on solving the BCS equation, revealed that this conclusion is too hasty and that the mechanism of pairing in nuclei with one filled shell is basically of a volume character, but the authors of [21] did not reject the mixed mechanism out of hand. Since the method of analysis in [20] was markedly different from that in [21], it is clear that the problem requires a more consistent consideration than in [17, 20].

As was indicated above, the refinement of the gap is one of the most important ingredients of the procedure for refining phenomenological parameters in nonmagic nuclei. The importance of this procedure is highlighted by the following.

The mass operator $\Sigma^{(2)}$ appearing in equations (5) can be represented in a form different from that in (6). Considering that, in the limit where there is no quasiparticle–phonon interaction, $\Sigma^{(2)}$ goes over to the conventional BCS gap and generalizing the corresponding approach in the theory of finite Fermi systems, we can symbolically write

$$-i\Sigma^{(2)}(\varepsilon) = \nabla F^{(2)}, \qquad (38)$$

where $\nabla$ is the irreducible block that cannot be broken down into parts connected by the corresponding Green's functions appearing in $F^{(2)}$ (30). When the quasiparticle–phonon interaction is taken into account in the $g^2$ approximation, we have [19, 20]

$$\nabla = W + gDg. \qquad (39)$$

The function $F^{(2)}$ must also be used in this approximation:

$$F^{(2)} \approx \tilde{F}^{(2)} + \tilde{F}^{(2)} M\tilde{G} + \tilde{G}^{(h)} M^{(h)} \tilde{F}^{(2)} - \tilde{F}^{(2)} M^{(1)} \tilde{F}^{(2)} + \tilde{G}^{(h)} M^{(2)} \tilde{G}. \qquad (40)$$

In (39), $gDg$ represents the exchange of a phonon between quasiparticles (phonon contribution in the *pp* channel), while $W$ is the corresponding residue of $\nabla$. By using relations (38), (39), and (40) and discarding terms of order $g^4$ and higher, we obtain

$$-i\Sigma^{(2)}(\varepsilon) = W\tilde{F}^{(2)} + gDg\tilde{F}^{(2)} + W\tilde{F}^{(2)} M\tilde{G} + \tilde{G}^{(h)} M^{(h)} \tilde{F}^{(2)} - \tilde{F}^{(2)} M^{(1)} \tilde{F}^{(2)} + \tilde{G}^{(h)} M^{(2)} \tilde{G}, \qquad (41)$$

which graphically corresponds to

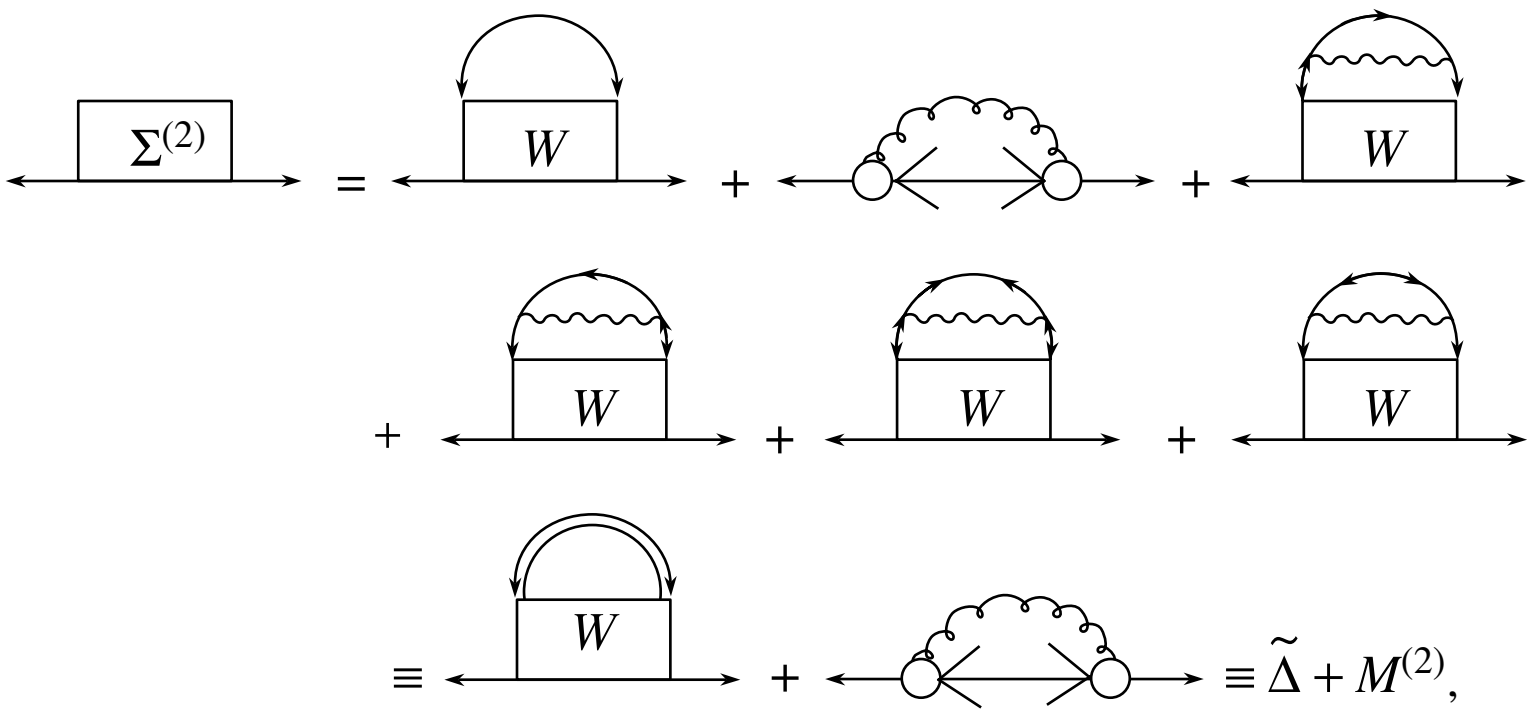

(42)

where the double line represents the Green's function $F^{(2)}$ (40) taken in the $g^2$ approximation. (It is the quantity $\Sigma^{(2)}$ that was considered in [22], where it was referred to as a gap.) But if the double line in (42) is taken to mean the total Green's function, we must sum $gDg$ terms over $g^2$ (this has not yet been done for the *pp* channel) by analogy with [9, 10]. Substituting relations (23)–(25) into (42), we obtain a set of equations that makes it possible to find $\Delta_{\lambda\eta}$ in the $g^2$ approximation.

It can be seen from equation (42) that, in the expression for the gap allowing for the quasiparticle–phonon interaction, the BCS mechanism appears through an unknown interaction $W$ different from the conventional phenomenological *pp* interaction. In addition, this expression involves rather complicated terms containing $W$ and $g^2$. In view of all the above, it is advisable to define $\Sigma^{(2)}$ according to (6) and to find $\tilde{\Delta}$ phenomenologically by following the proposed procedure of refining. From (42), it follows that, if $W$ is independent of energy, the quantity $\tilde{\Delta}$ is independent of energy as well, but it contains graphs with insertions featuring $g^2$. By determining $\tilde{\Delta}$ from our simple procedure of refining, we thereby avoid solving complicated equations for the gap with the unknown interaction $W$, but we then take explicitly into account only the pole part $M^{(2)}$ in the mass operator $\Sigma^{(2)}$. Quantitatively, this procedure, which makes it possible to sidestep the problem of doubly taking into account the quasiparticle–phonon interaction, is quite satisfactory because, from the quantity $\Delta_\lambda$, we extract the contribution of those phonons that are included explicitly in the subsequent calculations. This contribution is determined by $M_\lambda^{(1,2)}$. The greatest contribution is expected from lowly lying collective phonons, which are predominantly superficial [23], featuring only a slight volume component.

## 4. CALCULATION OF SINGLE-PARTICLE STRENGTHS IN $^{119}$Sn AND $^{121}$Sn

In this section, we present the first results obtained by calculating the distributions of single-particle strengths in $^{119}$Sn and $^{121}$Sn nuclei by formulas (26), (32), and (34). The basic results of these calculations are presented in Figs. 1–8 and in Table 3, along with available experimental data. Much attention is given to the procedure of refining and to the role of new terms in the mass operator.

### *4.1. Constructing Single-Particle Spectrum: Details of the Calculations*

Information about the single-particle spectrum is used as input data in many calculations. This information can be obtained from available experimental data on nucleon-separation energies and on the excitation energies of the odd neighboring nuclei. Having such data at our disposal and choosing appropriately mean-field potentials in the independent-particle model, we can construct the single-particle spectrum. It should be emphasized, however, that, because of considerable uncertainties in obtaining and processing relevant experimental data, such information is not always reliable for nonmagic nuclei.

Since pairing plays an important role for nonmagic nuclei, our arguments will be based on the model of independent Bogolyubov quasiparticles. The values of $E_\lambda = \sqrt{\varepsilon_\lambda^2 + \Delta_\lambda^2}$ can be taken directly from experimental data by using the excitation energies $E_\lambda = (E_{N+2} + E_N - 2E_{N+1}^{(\lambda)})/2$ of odd nuclei, where $E_{N+2}$ and $E_N$ are the experimental binding energies of ground-state even nuclei featuring $N + 2$ and $N$ particles, respectively, while $E_{N+1}^{(\lambda)}$ are the experimental binding energies of the single-quasiparticle states of the odd $(N + 1)$ nucleus [24]. Since $E_{N+1}^{(\lambda)} = E_{N+1} - E_{\text{exc}}^{\lambda}$, where $E_{N+1}$ is the experimental binding energy of the ground state of the odd $(N + 1)$ nucleus, and $E_{\text{exc}}^{\lambda}$ is the excitation energy of the odd nucleus with respect to its ground state, we can write $E_\lambda = \bar{\Delta} + E_{\text{exc}}^{\lambda}$, where $\bar{\Delta}$ is the odd–even mass difference.

Knowing the experimental values of $E_\lambda$, we can find the single-particle energies $\varepsilon_\lambda$ and the gap values $\Delta_\lambda$ by solving the equation for the gap under the condition requiring that the available experimental values of $E_\lambda$ be reproducible with the resulting values of $\Delta_\lambda$ and $\varepsilon_\lambda$. In order to achieve this, we employed the following iterative procedure:

(i) With the input set of single-particle energies and wave functions $(\varepsilon_\lambda^0, \varphi_\lambda^0)$ that is determined by solving the single-particle Schrödinger equation with the Woods–Saxon potential featuring conventional parameter values, the equation for the gap is iterated until the above condition is satisfied. (Previously, this condition, which automatically reproduces the odd–even mass difference as well, was commonly used in solving the equation for the gap.) In this way, we obtain the set $(\varepsilon_\lambda^1, \Delta_\lambda^1)$.

(ii) By slightly changing the potential-well depth, we find the mean-field potentials that reproduce the resulting $\varepsilon_\lambda^1$. After that, the procedure described in item (i) is applied to the set $(\varepsilon_\lambda^1, \varphi_\lambda^1)$. The entire procedure is repeated until the eventual phenomenological basis $(\varepsilon_\lambda, \varphi_\lambda)$ is obtained with the required accuracy.

**Table 1.** Single-particle energies $\varepsilon_\lambda$ and gap values $\Delta_\lambda$ for the $^{120}$Sn nucleus that were obtained from the experimental values of $E_\lambda$ [25] (the chemical potential value was taken to be $\mu = -6.73$ MeV)

| $\lambda$ | $\varepsilon_\lambda$, MeV | $\Delta_\lambda$, MeV | $E_\lambda^{\text{exp}}$, MeV |
|---|---|---|---|
| $1g9/2$ | −13.75 | 1.59 | |
| $2d5/2$ | −9.4 | 1.32 | |
| $1g7/2$ | −8.09 | 1.63 | 2.1 |
| $2d3/2$ | −6.82 | 1.32 | 1.32 |
| $3s1/2$ | −6.6 | 1.26 | 1.32 |
| $1h11/2$ | −6.3 | 1.47 | 1.41 |
| $2f7/2$ | −1.38 | 1.01 | |
| $3p3/2$ | −0.40 | 0.65 | |
| $3p1/2$ | 0.14 | 0.48 | |
| $1h9/2$ | 1.31 | 1.52 | |
| $2f5/2$ | 1.60 | 0.99 | |

In a given nucleus, there exist a few levels that can be treated as single-particle ones, and they naturally occur in the vicinity of the Fermi surface. [By single-particle states (levels), we mean here levels in the independent-particle model that are obtained in the Woods–Saxon potential.] For example, the $^{120}$Sn nucleus has four such levels ($2d3/2$, $3s1/2$, $1h11/2$, and $1g7/2$), and it is precisely their single-quasiparticle nature that serves as a basic point for choosing the input approximation in developing a single-particle scheme. For this nucleus, two iterations were required to a achieve a precision of 0.05 MeV in determining the energies of single-particle levels. The results of our calculations performed for the $^{120}$Sn nucleus are given in Table 1, which displays only 11 levels near the Fermi surface (actually, we took into account single-particle levels from the bottom of the potential well up to 20 MeV). (In the heading of this table, we quote the chemical-potential value of $\mu = -6.73$ MeV, which was obtained from particle-number-conservation condition used in the fitting procedure.)

For the $^{120}$Sn nucleus, phonons were computed on the basis of the theory of finite Fermi systems. In this calculation, the parameters of the Landau–Migdal forces [10, 26] were set to the known values of

$$f_{\text{in}} = -0.002, \quad f'_{\text{ex}} = 2.30, \quad f'_{\text{in}} = 0.76,$$

$$g = -0.05, \quad g' = 0.96, \quad C_0 = 300 \text{ MeV fm}^3.$$

The parameter $f_{\text{ex}}$ was adjusted in such a way as to obtain the experimental energy values of $\omega(2_1^+) = 1.17$ MeV and $\omega(3_1^-) = 2.40$ MeV for the collective lowly lying phonons; this resulted in $f_{\text{ex}}(2_1^+) = -2.795$ and $f_{\text{ex}}(3_1^-) = -3.4$. After that, all phonons of even and odd multipole orders were calculated with $f_{ex}(2_1^+)$ and $f_{ex}(3_1^-)$, respectively.

**Table 2.** Refined neutron single-particle energies $\tilde{\varepsilon}_\lambda$ and gap values $\tilde{\Delta}_\lambda$ for $^{120}$Sn

| $\lambda$ | $\varepsilon_\lambda$, MeV | $\tilde{\varepsilon}_\lambda$, MeV | | $\Delta_\lambda$, MeV | $\tilde{\Delta}_\lambda$, MeV, 21 phonons | $\gamma_\lambda$, % | |
|---|---|---|---|---|---|---|---|
| | | 21 phonons | 3 phonons | | | 21 phonons | 3 phonons |
| 1*g*9/2 | −7.02 | −7.67 | −6.87 | 1.62 | 0.94 | 42 | −3 |
| 2*d*5/2 | −2.67 | −3.69 | −3.41 | 1.35 | −0.23 | 117 | 85 |
| 1*g*7/2 | −1.36 | −3.27 | −2.54 | 1.58 | 0.76 | 52 | 33 |
| 2*d*3/2 | −0.09 | −0.17 | −0.24 | 1.36 | 1.09 | 20 | 18 |
| 3*s*1/2 | 0.13 | 0.37 | 0.20 | 1.27 | 0.89 | 26 | 16 |
| 1*h*11/2 | 0.42 | 0.86 | 0.91 | 1.62 | 1.41 | 13 | 9 |
| 1*f*7/2 | 5.35 | 5.61 | 5.42 | 1.11 | 1.39 | −25 | −37 |
| 3*p*3/2 | 6.32 | 7.43 | 7.66 | 0.70 | 0.53 | 24 | 4 |
| $\mu = -6.73$ MeV | | $\tilde{\mu} = -6.71$ MeV | $\tilde{\mu} = -6.82$ MeV | | | $\bar{\gamma} = 32$ | $\bar{\gamma} = 15$ |

In the present calculations, we did not study the convergence of the method in the number of phonons. It seems reasonable to take into account the phonons that are characterized by the highest degree of collectivity and which have energies not exceeding the neutron binding energy. We restricted ourselves to $2^+$, $3^-$, $4^+$, $5^-$, and $6^+$ phonons, for which the deformation parameter is greater than 0.03. There appeared to be 21 such phonons in the $^{120}$Sn nucleus, and they were used in all the calculations. However, some calculations employed only the three lowest phonons with the highest degree of collectivity.

The input data for our calculations also included the values of $\varepsilon_\lambda$ and $\Delta_\lambda$ as determined according to the procedure described above. In solving the BCS equation, we relied on the method used in the theory of finite Fermi systems to go over to the amplitude of particle–particle interaction [11, 21], $\Gamma^\xi = C_0/\ln(c_p/\xi)$, where $\xi$ is the parameter that cuts off the relevant sums over single-particle states in such a way that these sums are taken over the interval $\xi - \mu < \varepsilon_\lambda < \xi + \mu$, $\mu$ being the chemical potential. The constant value of $c_p = 1.36$ was chosen in such a way as to ensure the agreement of $\bar{\Delta}$ with the experimental odd–even mass difference.

### 4.2. Refining Single-Particle Spectrum and Gap Values

The refined values $\tilde{\varepsilon}_\lambda$ and $\tilde{\Delta}_\lambda$ were determined by solving the set of nonlinear equations (26) with the mass operators $M^i$ (see Appendix). In constructing the relevant solution, it was required that the number of refined quasiparticles be equal to the number of particles that was used to determine the chemical potential $\tilde{\mu}$. As was indicated above (see Section 4.1), only four "good" single-quasiparticle levels could be singled out from experimental data for the $^{120}$Sn nucleus. Since these experimentally observable single-quasiparticle levels were used as a basis for obtaining the phenomenological values of $\varepsilon_\lambda$ and $\Delta_\lambda$, it seems natural to refine their energies first. It is more reasonable, however, to perform refining for a greater number of levels for the following reasons: (a) Both for Bogolyubov quasiparticles and for refined quasiparticles, their number must be equal to the number of intranuclear nucleons. (b) For either type of quasiparticles, the occupation numbers for levels lying far from the Fermi surface are close to 0 (above the Fermi surface) or 1 (below it); therefore, the occupation numbers will change upon refining for levels in the vicinity of the Fermi surface, so that it is desirable to include at least these levels in the calculations. The results of these calculations are displayed in Table 2, which also lists the values of

$$\gamma_\lambda = \frac{\Delta_\lambda - \tilde{\Delta}_\lambda}{\Delta_\lambda}, \quad \bar{\gamma} = \frac{\sum_\lambda \gamma_\lambda(2j+1)}{\sum_\lambda (2j+1)}, \tag{43}$$

which represent the contribution of the quasiparticle–phonon pairing mechanism caused by the retarded *pp* interaction—that is, the explicit contribution from the mass operators $M^{(1,2)}$.

The quantities $\tilde{\varepsilon}_\lambda$ and $\tilde{\Delta}_\lambda$ are unobservable, $\tilde{\Delta}_\lambda$ determining the contribution of the mechanism that is analogous to the BCS mechanism as given by (42), but which is not identical to it. Table 2 presents the results of our calculations that took account of the lowest three phonons ($2_1^+$, $3_1^-$, $4_1^+$), which are the most collective ones, and of 21 phonons. It can be seen from this table that the above three phonons are responsible for more than half of the contribution to the refining of the levels close to the Fermi surface. The $\lambda$ dependence of the quantities $\gamma_\lambda$ is significant, which is determined by the strength of the coupling of this single-particle state to collective states, especially to the first lowly lying ones. The averaged value $\bar{\gamma}$ as defined in (43) and as evaluated by using the levels presented in Table 2 amounts to about 32%. The role of refining is illustrated in Fig. 2, which displays the distributions of the single-particle strength for the $2d5/2$ level that were calculated with and without refining.

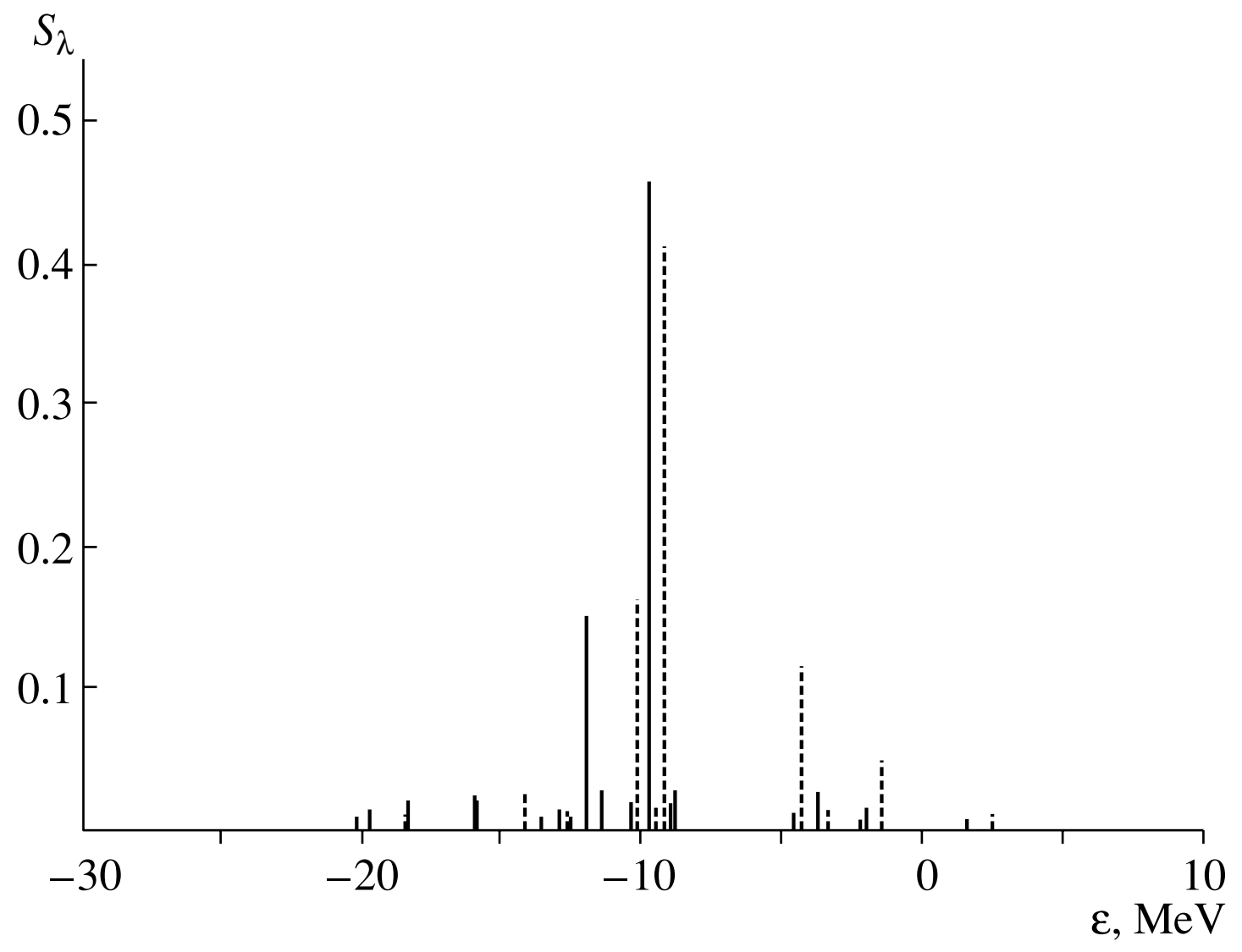

**Fig. 2.** Distributions of the single-particle strength of the $2d5/2$ neutron state with respect to $\varepsilon = \mu \pm (\bar{\Delta} \pm E_{exc})$ according to the calculation with 21 phonons for $^{119}$Sn (to the left of the Fermi energy of –6.73 MeV) and $^{121}$Sn (to the right of the Fermi energy): (solid lines) results obtained with allowance for refinement of $\varepsilon_\lambda$ and $\Delta_\lambda$ and (dashed lines) results obtained without refinements.

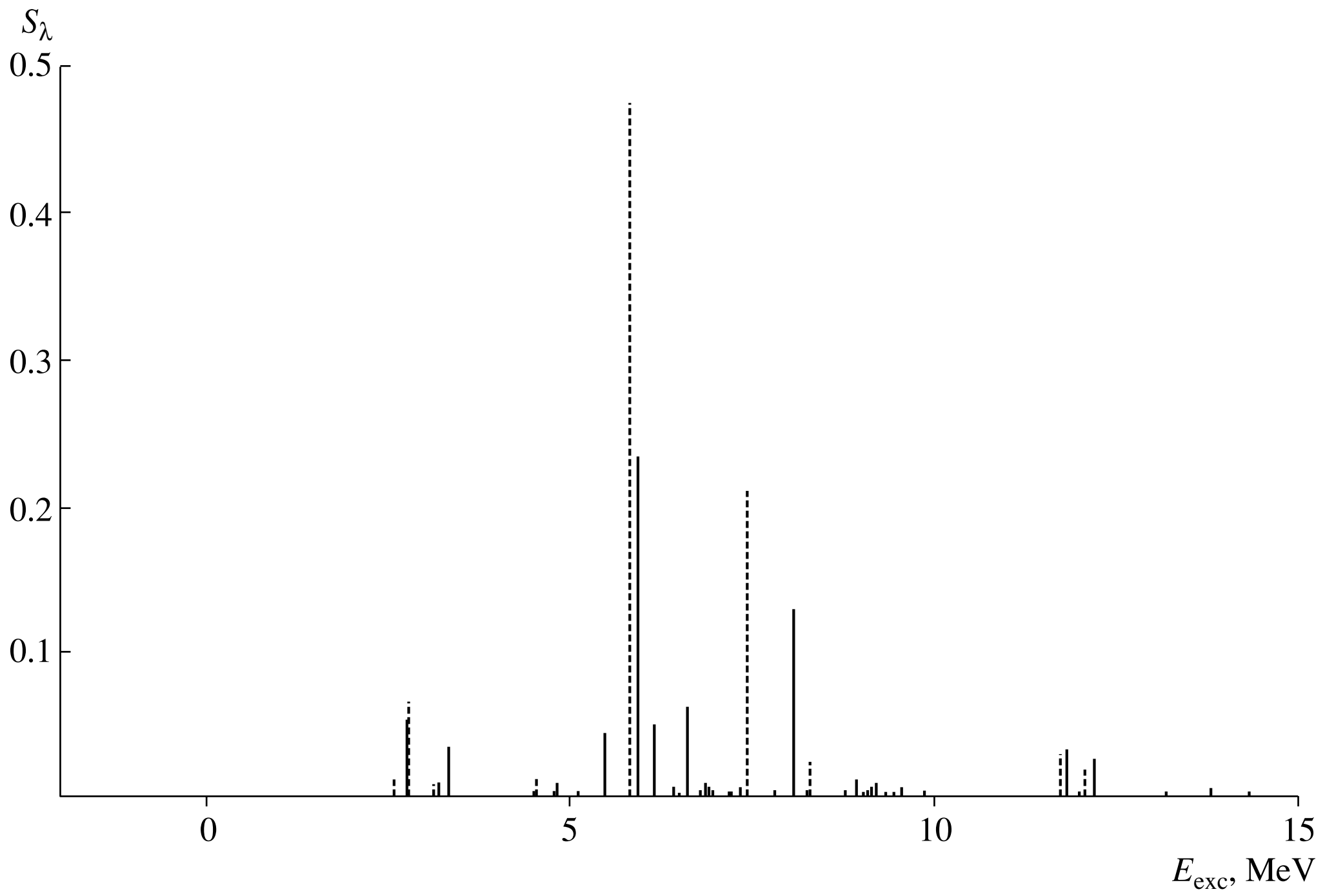

**Fig. 3.** Distributions of the single-particle strength of the $1g9/2$ neutron state in $^{119}$Sn with respect to the excitation energy of the nucleus: (solid lines) results of the calculation with 21 phonons and (dashed lines) results of the calculation with 3 phonons.

The calculations revealed that the role of refining considerably increases for single-particle states separated from the Fermi surface by 5–7 MeV. In other words, the role of refining becomes less pronounced as the strength distribution becomes more similar to that for a Bogolyubov quasiparticle in a given state. For example, the role of refining proved insignificant for $3s1/2$ and $1h11/2$ states.

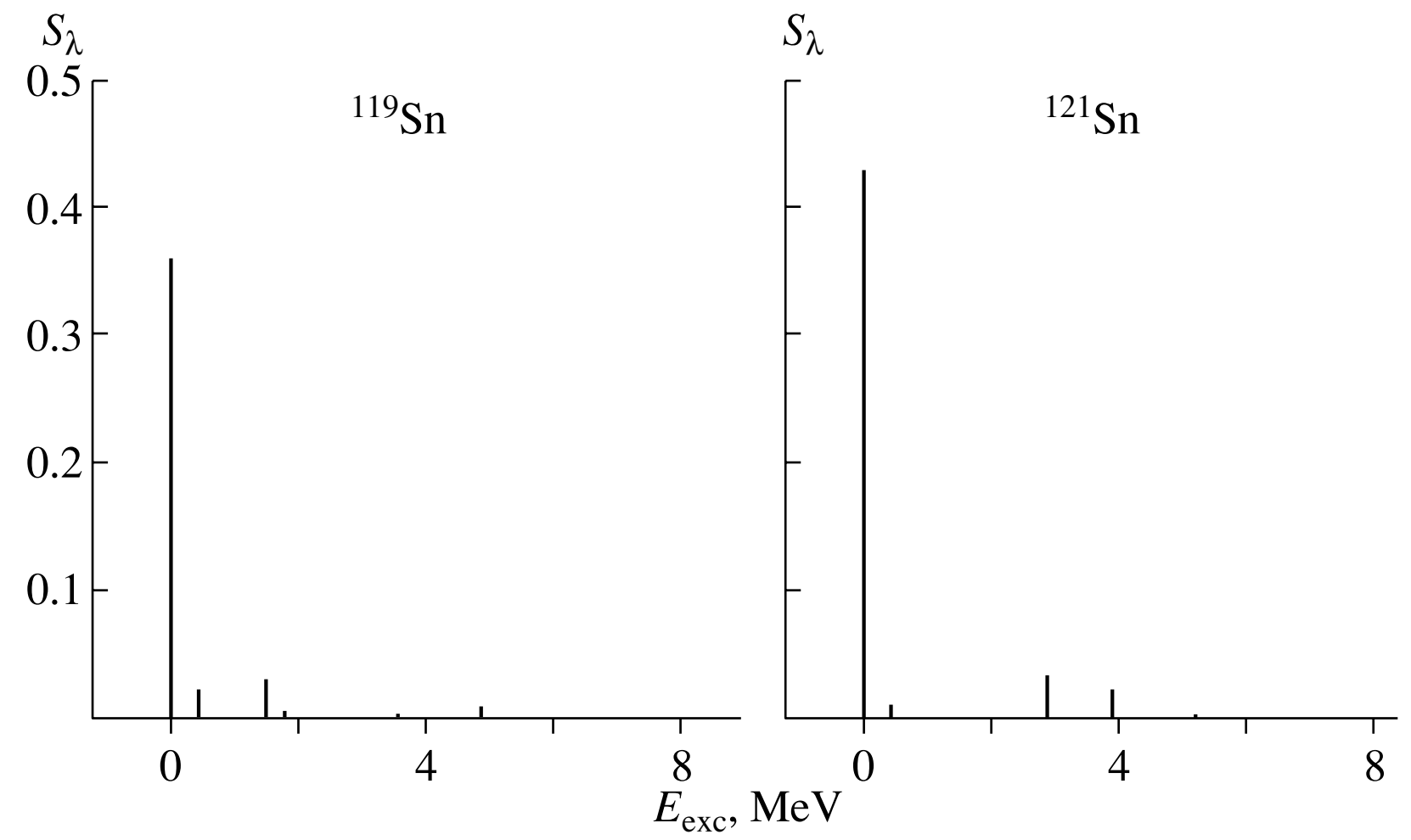

**Fig. 4.** Distributions of the single-particle strength of the $3s1/2$ neutron state in $^{119}$Sn and $^{121}$Sn.

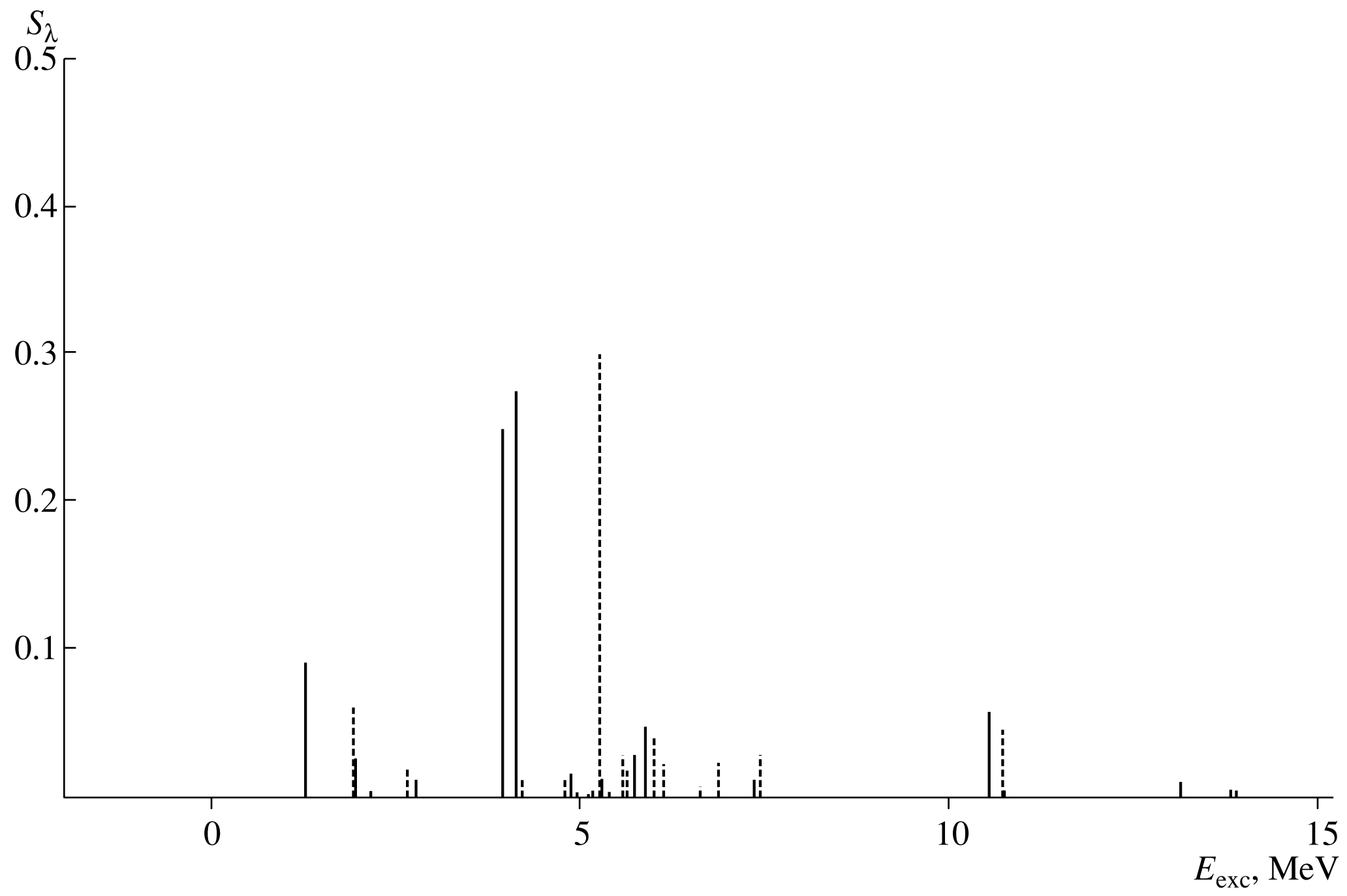

**Fig. 5.** Distributions of the single-particle strength of the $2f\,7/2$ neutron state in $^{121}$Sn: (solid lines) results of the full calculation and (dashed lines) results of the calculation disregarding the mass operators $M^{(1, 2)}$.

### *4.3. Distributions of Single-Particle Strengths in $^{119}$Sn and $^{121}$Sn*

A feature peculiar to nonmagic nuclei is that, because of pairing, a single-particle level of the even–even $N$ nucleus is observed with a sizable probability in both neighboring odd ($N \pm 1$) nuclei (see Figs. 4 and 8). In the model of independent quasiparticles, these probabilities are proportional to the Bogolyubov coefficients $u_2^2$ and $v_2^2$.

The dependences of the results on the number of phonons taken into account and on the inclusion of the additional terms $M^{(1, 2)}$ are illustrated in Figs. 3 and 4 and Figs. 5–8, respectively. For states corresponding to single-particle levels close to the Fermi surface, the calculations with twenty-one and three phonons lead to virtually identical results. This is exemplified by the strength distributions for the $3s1/2$ states in $^{119}$Sn and $^{121}$Sn nuclei (see Fig. 4, where the distributions obtained with three phonons are not shown because of the smallness of the effect). A similar pattern is observed for $2d3/2$ and $1h11/2$ (Fig. 8) states close to the Fermi surface. Thus, we conclude that, for these states, the main contribution to the strength distributions from quasiparticle–phonon interaction is due to the three lowest phonons ($2_1^+$, $3_1^-$, and $4_1^+$ ones). However, the inclusion of a greater number of phonons has a pronounced effect for single-particle states remote from the Fermi surface (see Fig. 3).

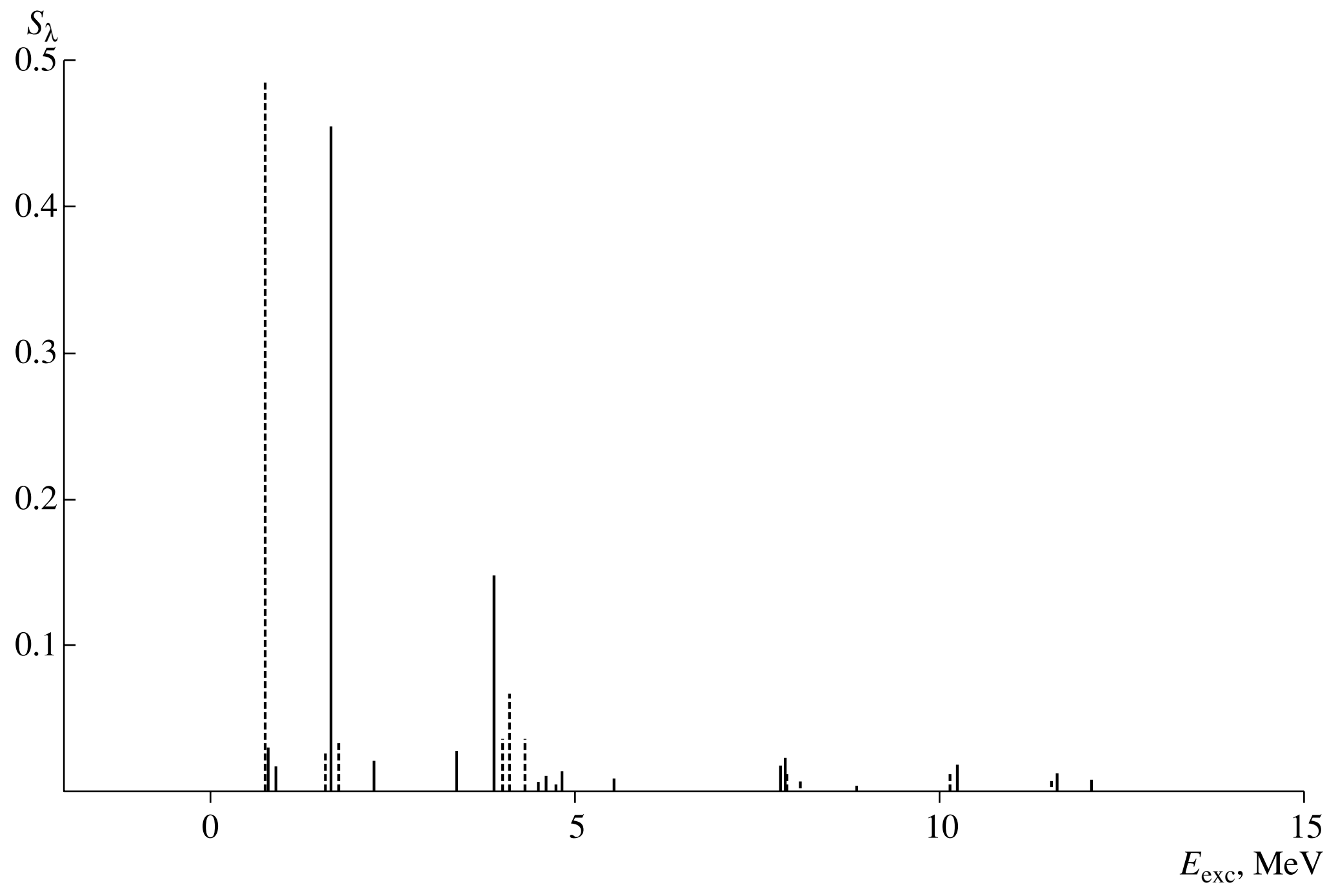

**Fig. 6.** As in Fig. 5, but for the $2d5/2$ state in $^{119}$Sn.

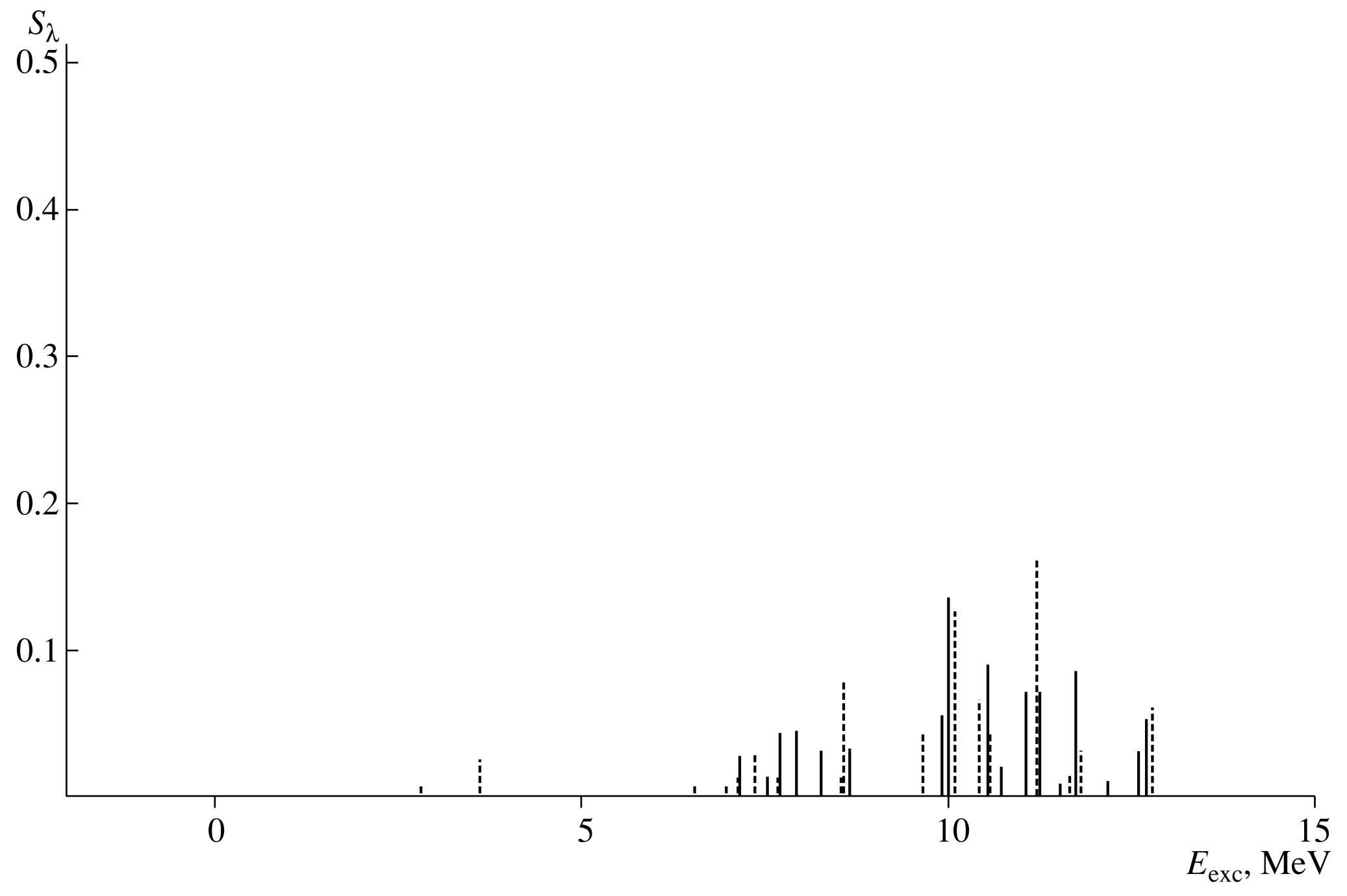

**Fig. 7.** As in Fig. 5, but for the $1f5/2$ state in $^{119}$Sn.

Figures 5–8 illustrate the effect of the mass operators $M^{(1,2)}$ on the distributions of single-particle

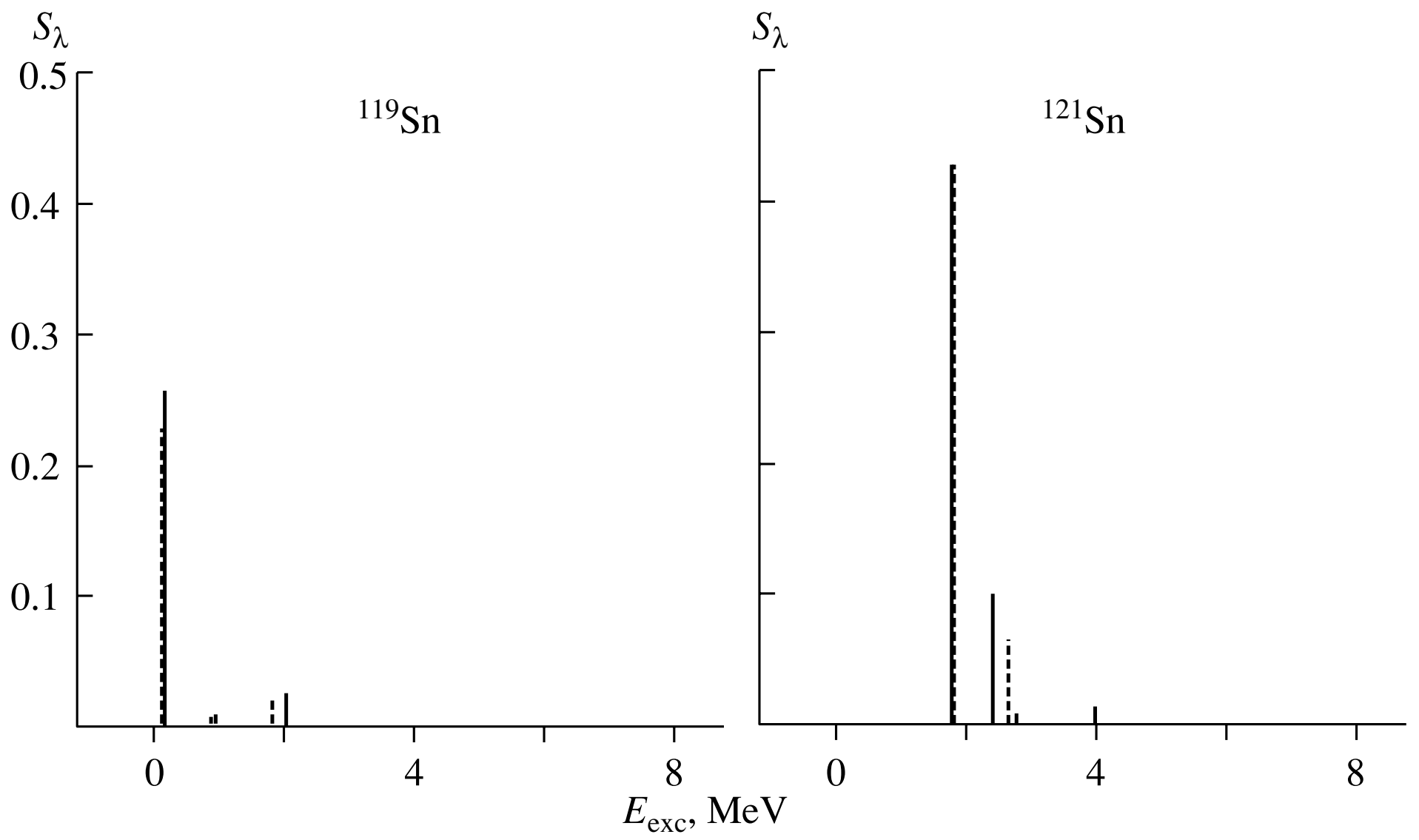

**Fig. 8.** As in Fig. 5, but for the 1*h*11/2 state in $^{119}$Sn and $^{121}$Sn.

strengths. The pattern here is similar to the preceding one in that these mass operators exert markedly different effects for states remote to the Fermi surface and states close to it. Specifically, the inclusion of $M^{(1,2)}$ leads to sizable changes (see Figs. 5–7) in the strength distributions for states remote from the Fermi surface; for states occurring in the immediate vicinity of the Fermi surface, the effect of this retarded interaction is much weaker (Fig. 8), and we can additionally note here smaller values of the main spectroscopic factors when $M^{(1,2)}$ is disregarded. Thus, we can state that, for states close to the Fermi surface, pairing is dominated by the BCS mechanism. The pairing mechanism described by the $M^{(1,2)}$ terms comes into play for states remote from the Fermi surface. However, the integrated features do not show strong changes in the energy regions being considered.

In Table 3, the results of the calculations are contrasted against available experimental data for the neutron states corresponding to ten single-particle levels on both sides of the Fermi surface. The experimental data were borrowed from [25]. The results of the calculations are seen to be in fairly good agreement with these data.

In order to test our computational procedure, we used the sum rule for the spreading of a given shell state

**Table 3.** Energies and spectroscopic factors of neutron states in (upper lines) $^{119}$Sn and (lower lines) $^{121}$Sn

| $\lambda$ | $E_{exc}$, MeV | | $S_\lambda$ | | $v_\lambda^2$ / $u_\lambda^2$ | $\lambda$ | $E_{exc}$, MeV | | $S_\lambda$ | | $v_\lambda^2$ / $u_\lambda^2$ |
|---|---|---|---|---|---|---|---|---|---|---|---|
| | expt. | theor. | expt. | theor. | | | expt. | theor. | expt. | theor. | |
| 2*p** | 7.10 | 7.50 | 2.42 (4.3–10.2) | 2.13 | 1 | 3*s*1/2 | 0 | ≈0 | 0.26; 0.32 | 0.36 | 0.45 |
| | | | | | 0 | | 0.06 | ≈0 | 0.3 | 0.43 | 0.55 |
| 1*g*9/2 | 5.41 | 5.90 | 0.53 (3.9–6.5) | 0.43 | ≈1 | 1*h*11/2 | 0.09 | 0.18 | 0.29 | 0.26 | 0.37 |
| | | | | | ≈0 | | 0.06 | 0.17 | 0.49 | 0.43 | 0.63 |
| 2*d*5/2 | 1.10 | 1.69 | 0.43 | 0.46 | 0.95 | 2*f*7/2 | | | | | ≈0 |
| | 1.14 | 1.41 | 0.11 (0.0–2.0) | 0.04 | 0.05 | | 2.83 | 3.11 | 0.35 (0.0–4.0) | 0.39 | ≈1 |
| 1*g*7/2 | 0.79 | 1.81 | 0.75; 0.6 | 0.66 (0.0–3.0) | 0.83 | 3*p*3/2 | | | | | ≈0 |
| | 2.85 | 1.23 | 0.15 (0.0–3.5) | 0.04 | 0.17 | | 3.73 | 4.97 | 0.54 (0.0–5.0) | 0.42 | 1 |
| 2*d*3/2 | 0.02 | 0.01 | 0.4; 0.45 | 0.4 | 0.52 | 1*h*9/2 | | | | | 0 |
| | 0 | ≈0 | 0.44; 0.65 | 0.35 | 0.48 | + 1*i*13/2* | 7.5 | 7.47 | ≈24 | 19.92 | 1 |

Note: In parentheses, we indicate intervals where the spectroscopic factors were summed for states without a distinct dominant peak. The energies $E_{exc}$ correspond to the position of the dominant peak or to the weighted-mean energy if no such peak can be found.

* The quantities $S_\lambda(2j + 1)$ are presented.

$\lambda$: $\sum_\eta S_{\eta\lambda} = 1$. In a broad energy interval, this sum rule was exhausted completely; in the intervals shown in the figures, our results saturated more than 90% of it.

## 5. CONCLUSION

A realistic model for calculating the distributions of single-particle strengths in nuclei has been formulated and implemented. The model describes excited states in odd nonmagic nuclei. It is applicable, above all, to spherical nuclei with one unfilled shell. [A simple case including composite (quasiparticle⊗phonon) configurations has been considered, but the developed approach makes it possible to take straightforwardly into account more complicated configurations of the (quasiparticle⊗two phonons type).] It was desirable to develop such a model because experimental data on nonmagic nuclei—and these data are becoming much vaster and, what is more important, more precise—require a thorough explanation. Here, we have achieved reasonable agreement with available data on $^{119}$Sn and $^{121}$Sn nuclei.

The following three features of our model give sufficient grounds to believe that it will prove viable. First, it consistently employs the $g^2$ approximation for the ground and excited states both in the *pp* and in the *ph* channel. It is also important that ground-state correlations have been fully taken into account within the approximations used. Second, the procedure of refining the phenomenological values of $\varepsilon_\lambda$ and $\Delta_\lambda$, which seems to be especially important for the gap and, hence, for understanding the pairing mechanism, provides a quantitatively reasonable phenomenological approach to the problem: (i) There is no need to find a new *pp* interaction and to solve a new, much more complicated, equation for the gap—a simpler way consists in obtaining the refined values $\tilde{\varepsilon}_\lambda$ and $\tilde{\Delta}_\lambda$ in accordance with the recipe described in Subsection 2.2. (ii) From the quantities $\varepsilon_\lambda$ and $\Delta_\lambda$, we extract the contribution of those phonons that are taken into account in the ensuing calculations. Third, an auxiliary procedure for extracting single-particle energies $\varepsilon_\lambda$ from the observed quasiparticle energies $E_\lambda$ of nonmagic nuclei was implemented.

The feature that distinguishes our approach to nonmagic nuclei from what is usually done for magic nuclei is that, in addition to invoking the idea of Cooper pairing and Bogolyubov quasiparticles, we have supplemented the anomalous mass operators with the pole terms $M^{(1)}$ and $M^{(2)}$, which give rise to a retarded *pp* interaction induced by phonon exchange. These quantities have been consistently taken into account in describing both the ground state (refinement of the gap and of single-particle energies) and excited states. Our refinement of the gap proved to be greatly dependent on the single-particle state; its averaged value for $^{120}$Sn is about 32% of the experimental value. The inclusion of the quantities $M^{(1)}$ and $M^{(2)}$ proved necessary, at least for states remote from the Fermi surface.

## ACKNOWLEDGMENTS

We are grateful to V.I. Tselyaev for stimulating discussions.

This work was supported in part by the Russian Foundation for Basic Research (project no. 96-02-17250).

## APPENDIX

Let us consider the mass operators $M^i$ in the $\lambda$ representation. In the diagonal approximation in this index, we have

$$M_1(\varepsilon) = M_1^{(h)}(-\varepsilon)$$

$$= \sum_{s,2} (g_{12}^s)^2 \left( \frac{\tilde{u}_2^2}{\varepsilon - \tilde{E}_2 - \omega_s + i\gamma} + \frac{\tilde{v}_2^2}{\varepsilon + \tilde{E}_2 + \omega_s - i\gamma} \right),$$

$$M_1^{(1)}(\varepsilon) = M_1^{(2)}(\varepsilon)$$

$$= -\sum_{s,2} (g_{12}^s)^2 \frac{\tilde{\Delta}^2}{2\tilde{E}_2} \left( \frac{1}{\varepsilon - \tilde{E}_2 - \omega_s + i\gamma} - \frac{1}{\varepsilon + \tilde{E}_2 + \omega_s - i\gamma} \right),$$

where the subscript 1 stands for the set of quantum numbers $1 \equiv n_1, j_1, l_1$, and $m_1$; $\tilde{E}_2 = \sqrt{\tilde{\varepsilon}_2^2 + \tilde{\Delta}_2^2}$; and $g^s$ is the amplitude of the production of a phonon with energy $\omega_s$.

## REFERENCES

1. Soloviev, V.G., *Theory of Complex Nuclei*, Oxford: Pergamon, 1976.
2. Solov'ev, V.G., *Theory of Atomic Nuclei*, Bristol: IOP, 1992.
3. Gales, S., Stoyanov, Ch., and Vdovin, A.I., *Phys. Rep.*, 1988, vol. 166, p. 125.
4. Bohr, A. and Mottelson, B.R., *Nuclear Structure*, vol. 2: *Nuclear Deformations*, New York: Benjamin, 1975.
5. Bertsch, G.F., Bortignon, P.F., Broglia, R.A., and Dasso, C.H., *Phys. Lett. B*, 1979, vol. 80, p. 161.
6. Bertsch, G.F., Bortignon, P.F., and Broglia, R.A., *Rev. Mod. Phys.*, 1983, vol. 55, p. 287.
7. Kamerdzhiev, S.P., *Pis'ma Zh. Eksp. Teor. Fiz.*, 1979, vol. 30, p. 532.
8. Kamerdzhiev, S.P., *Yad. Fiz.*, 1983, vol. 38, p. 316.
9. Kamerdzhiev, S.P. and Tkachev, V.M., *Yad. Fiz.*, 1986, vol. 43, p. 1426.
10. Kamerdzhiev, S.P., Tertychnyi, G.Ya., and Tselyaev, V.I., *Fiz. Elem. Chastits At. Yadra*, 1997, vol. 28, p. 333.
11. Migdal, A.B., *Theory of Finite Fermi Systems and Applications to Atomic Nuclei*, New York: Interscience, 1967.
12. Ring, P. and Werner, E., *Nucl. Phys. A*, 1973, vol. 211, p. 198.

13. Eliashberg, G.M., *Zh. Eksp. Teor. Fiz.*, 1960, vol. 38, p. 966.
14. Landau, L.D. and Lifshitz, E.M., *Statistical Physics*, Oxford: Pergamon, 1980, part 2.
15. Kamerdzhiev, S.P., *Yad. Fiz.*, 1997, vol. 60, p. 572 [*Phys. At. Nucl.* (Engl. Transl.), vol. 60, p. 497].
16. Belyaev, S.T. and Zelevinskii, V.G., *Yad. Fiz.*, 1965, vol. 1, p. 13.
17. Kadmensky, S.G. and Luk'yanovich, P.A., *Yad. Fiz.*, 1989, vol. 49, p. 384.
18. van Neck, D., *PhD Dissertation*, Gent: Univ. of Gent, 1992; van Der Sluys V., van Neck, D., Waroquier, M., and Ryckebusch, J., *Nucl. Phys. A*, 1993, vol. 551, p. 210.
19. Hahne, F.J.W., Heiss, W.D., and Engelbrecht, C.A., *Ann Phys.* (N.Y.), 1977, vol. 104, p. 251; *Phys. Lett. B*, 1977, vol. 66, p. 218; Heiss, W.D., Engelbrecht, C.A., and Hahne, F.J.W., *Nucl. Phys. A*, 1977, vol. 289, p. 386.
20. Kadmensky, S.G., Luk'yanovich, G.A., Remezov, Yu.I., and Furman, V.I., *Yad. Fiz.*, 1987, vol. 45, p. 942.
21. Zverev, M.V. and Sapershtein, E.E., *Yad. Fiz.*, 1985, vol. 42, p. 1082.
22. Kamerdzhiev, S.P. and Avdeenkov, A.V., *XLVII Mezhdunar. soveshch. po yadernori spektroskopii i strukture atomnogo yadra* (Proc. Int. Meeting on Nuclear Spectroscopy and Nuclear Structure), St. Petersburg, 1997, p. 244.
23. Platonov, A.P. and Sapershtein, E.E., *Yad. Fiz.*, 1987, vol. 46, p. 437.
24. Krainov, V.P., *Lektsii po mikroskopicheskoi teorii yadra* (Lectures on the Microscopic Theory of the Nucleus), Moscow: Atomizdat, 1973.
25. *Nucl. Data Sheets*, 1992, vol. 67, no. 2; 1991, vol. 64, no. 2; Wapstra, H. and Audi, G., *Nucl. Phys. A*, 1985, vol. 432, p. 55.
26. Smirnov, A.V., Tolokonnikov, S.V., and Fayans, S.A, *Preprint of Inst. of Atomic Energy*, Moscow, 1986, no. IAE-4281/2; *Yad. Fiz.*, 1988, vol. 48, p. 1661; Borzov, I.N., Tolokonnikov, S.V., and Fayans, S.A., *Yad. Fiz.*, 1984, vol. 40, p. 1151.

*Translated by A. Isaakyan*